\documentclass[useAMS,usenatbib]{mn2e}

\usepackage{graphicx}
\usepackage{rotating}
\usepackage{pdflscape}
\usepackage{amssymb}
\usepackage{color}
\usepackage{url}

\title[Atmospheric Parameters of 169 F, G, K and M-type Stars in the Kepler Field]{Atmospheric Parameters of 169 F, G, K and M-type Stars in the {\it Kepler} Field\thanks{
Based on observations acquired at the Canada-France-Hawaii Telescope (CFHT) which is operated by the National Research Council of Canada, the Institut National des 
Sciences de l'Univers of the Centre National de la Recherche Scientifique of France, and the University of Hawaii, the Telescope Bernard Lyot (USR5026) operated by 
the Observatoire Midi-Pyr\'en\'ees (Universit\'e de Toulouse and the Institut National des Science de l'Univers of the Centre National de la Recherche Scientifique 
of France), the Nordic Optical Telescope, operated jointly by Denmark, Finland, Iceland, Norway, and Sweden, and with the Mercator telescope, operated by the 
Flemish Community, both located on the island of La Palma at the Spanish Observatorio del Roque de los Muchachos of the Instituto de Astrof\'{\i}sica de Canarias, 
and the M.G.~Fracastoro station of the INAF - Osservatorio Astrofisico di Catania, Italy. The Mercator observations were obtained with the HERMES spectrograph, 
which is supported by the Fund for Scientific Research of Flanders (FWO), Belgium, the Research Council of K.U.Leuven, Belgium, the Fonds National de la Recherche 
Scientifique (FNRS), Belgium, the Royal Observatory of Belgium, the Observatoire de Gen\`eve, Switzerland and the Th\"uringer Landessternwarte Tautenburg, Germany.
}}
\author[J.\,Molenda-\.Zakowicz, et al.]{J.\,Molenda-\.Zakowicz$^{1}$, S.G.\,Sousa$^{2}$, A.\,Frasca$^{3}$, K.\,Uytterhoeven$^{4,5}$, M.\,Briquet$^{6,7}$, \and
H.\,Van~Winckel$^{7}$, D.\,Drobek$^{1}$, E.\,Niemczura$^{1}$, P.\,Lampens$^{8}$, J.\,Lykke$^{9}$, S.\,Bloemen$^{7}$, \and J.F.\,Gameiro$^{2,10}$, C.\,Jean$^{7}$, 
D.\,Volpi$^{8}$, N.\,Gorlova$^{7}$, A.\,Mortier$^{2,10}$, M.\,Tsantaki$^{2,10}$, \and G.\,Raskin$^{7}$\\
$^{1}$ Instytut Astronomiczny Uniwersytetu Wroc{\l}awskiego, ul.\ Kopernika 11, 51-622 Wroc{\l}aw, Poland, E-mail: molenda@astro.uni.wroc.pl\\
$^{2}$ Centro de Astrof{\'i}sica, Universidade do Porto, Rua das Estrelas, 4150-762 Porto, Portugal\\
$^{3}$ INAF, Osservatorio Astrofisico di Catania, via S. Sofia, 78, 95123 Catania, Italy\\
$^{4}$ Instituto de Astrof\'isica de Canarias, 38200 La Laguna, Tenerife, Spain\\
$^{5}$ Departamento de Astrof\'isica, Universidad de La Laguna, 28206 La Laguna, Tenerife, Spain\\
$^{6}$ Institut d'Astrophysique et de G\'eophysique, Universit\'e de Li\`ege, All\'ee du 6 Ao\^ut 17, B\^at B5c, 4000 Li\`ege, Belgium\\
$^{7}$ Instituut voor Sterrenkunde, KU Leuven, Celestijnenlaan 200D, 3001 Leuven, Belgium\\
$^{8}$ Koninklijke Sterrenwacht van Belgi\"e, Ringlaan 3, 1180 Brussel, Belgium\\
$^{9}$Nordic Optical Telescope, 38700 Santa Cruz de La Palma, Spain\\
$^{10}$Departamento de F\'isica e Astronomia, Faculdade de Ci\^encias, Universidade do Porto, Portugal\\
}
\begin{document}

\date{Accepted 1988 December 15. Received 1988 December 14; in original form 1988 October 11}

\pagerange{\pageref{firstpage}--\pageref{lastpage}} \pubyear{2002}

\maketitle

\label{firstpage}

\begin{abstract}
The asteroseismic and planetary studies, like all research related to stars, need precise and accurate stellar atmospheric parameters as input. We aim at deriving 
the effective temperature ($T_{\rm eff}$), the surface gravity ($\log g$), the metallicity ([Fe/H]), the projected rotational velocity ($v\sin i$) and the MK type 
for~169 F, G, K, and M-type {\it Kepler} targets which were observed spectroscopically from the ground with five different instruments. We use two different 
spectroscopic methods to analyse 189 high-resolution, high-signal-to-noise spectra acquired for the 169 stars. For 67 stars, the spectroscopic atmospheric parameters 
are derived for the first time. KIC~9693187 and 11179629 are discovered to be double-lined spectroscopic binary systems. The results obtained for those stars for which 
independent determinations of the atmospheric parameters are available in the literature are used for a~comparative analysis. As a result, we show that for solar-type 
stars the accuracy of present determinations of atmospheric parameters is $\pm$~150~K in $T_{\rm eff}$, $\pm$~0.15~dex in [Fe/H], and $\pm$~0.3~dex in $\log g$. 
Finally, we confirm that the curve-of-growth analysis and the method of spectral synthesis yield systematically different atmospheric parameters when they are applied 
to stars hotter than 6,000~K. 
\end{abstract}


\section{Introduction}
Since March 2009, the 105 deg$^2$ field located in between the constellations of Cygnus and Lyra has been continuously monitored by the NASA space mission {\it 
Kepler} \citep{borucki2003, koch2010}. The effective temperature ($T_{\rm eff}$), surface gravity ($\log g$), and metallicity ([Fe/H]) of stars in the {\it 
Kepler} field were derived from the Sloan $griz$ photometry and are provided in the {\it Kepler} Input Catalog \citep[KIC,][]{brown2011}. KIC was created with 
the aim of providing a distinction between main-sequence stars and giants in the temperature range from 4,500 to 6,500~K. Within that range, the nominal 
precision of the values of $T_{\rm eff}$ and $\log g$ in KIC is 200~K and 0.5~dex, respectively. For hotter and cooler stars, the values of $T_{\rm eff}$ and $\log g$ 
in KIC become imprecise, while the estimates of [Fe/H] are poor in general \citep{brown2011}. 

This situation is very unfortunate for asteroseismic and planetary studies which require precise and accurate atmospheric parameters of stars to produce reliable 
results \citep[see, e.g.,][]{stello2009, creevey2012}. Therefore, ground-based follow-up observations aiming at deriving the values of $T_{\rm eff}$, $\log g$, and 
[Fe/H] are essential for further investigation of the {\it Kepler} targets. Such programmes started well before the {\it Kepler} satellite was 
launched \citep[see][]{molenda2007} and then, after the successful launch of the mission, were continued in the framework of the {\it Kepler} Asteroseismic Science 
Consortium\footnote{\url{http://astro.phys.au.dk/KASC}} (KASC) as a series of coordinated observing programmes for systematic spectroscopic and photometric observations
\citep[see][]{Uytterhoeven2010a, Uytterhoeven2010b}.

The first results of those proposals have been published by \citet{molenda2010, molenda2011, frasca2011, frohlich2012, bruntt2012, thygesen2012}. In this paper, we report 
the results of spectroscopic analysis of a next subset of F, G, K, and M-type stars. In Sect.~\ref{targets}, we outline the method of selecting targets. In 
Sect.~\ref{observations}, we provide information about the instruments and data acquisition, reduction and calibration. Our methods of analysis are described in 
Sect.~\ref{tgmet}. In Sect.~\ref{results}, the atmospheric parameters are provided and compared with other determinations reported in the literature. Sect.~\ref{discussion} 
contains a discussion of the accuracy of our results and the accuracy of the determinations of the atmospheric parameters of the solar-type stars in general. 
Sect.~\ref{summary} provides a summary.

\section{Target selection}
\label{targets}

Targets for each instrument were selected slightly differently. Those selected for the FIES spectrograph at the Nordic Optical Telescope (the principal investigator: KU)
and the HERMES spectrograph at the Mercator telescope (the principal investigator: MB and EN) included solar-like p-mode oscillators, $\gamma$~Dor, $\delta$~Sct, and 
$\beta$~Cep--type stars, and stars in the open clusters NGC~6811 and NGC~6819 requested for observations by the KASC community. In this paper, we analyse the 
F, G, K, and M-type stars observed with FIES (ten stars) and HERMES (twenty stars). The results obtained for early-type stars will be presented by 
Niemczura et al.\,and Catanzaro et al.\,(in prep.) 
When prioritising targets in those two proposals, more weight was given to stars that showed a particular interesting variable signal in the {\it Kepler} 
light-curves and hence promise to be the best targets for a comprehensive asteroseismic study, and to stars that were of interest to different KASC working groups. Brightness of the 
stars was another important selection factor. Since we made use of medium- and high-resolution spectrographs at 1-m and 3-m-class telescopes, we were limited to stars 
brighter than about $V=13$~mag. The final list of targets observed with HERMES included stars falling into the magnitude range of $10>V>8$~mag, while those observed with 
FIES, into the range of $11.5>V>7$~mag. 

The 18 stars which were observed with the FRESCO spectrograph at the 91-cm telescope at INAF-Osservatorio Astrofisico di Catania (INAF-OACt, the principal 
investigator: JM-\.Z) were selected from faint ($11>V>8$~mag), late-type ($1.7>B-V >0.5$~mag), close (the parallax $\pi >20$~mas) stars in the Tycho catalogue 
\citep{hog2000} which are optical counterparts of X-ray sources in the ROSAT All-Sky Survey Catalogue \citep[see][]{guillout1999}. These stars were proposed for 
{\it Kepler} asteroseismic targets and for the follow-up ground-based observations by AF in the first call for proposals announced by KASC. 

Our list of programme stars includes also 91 {\it Kepler} targets which were observed with the ESPaDOnS spectrograph at the Canada-France-Hawaii Telescope (the principal 
investigator: Claude Catala) and 50 stars observed with the NARVAL spectrograph at the Bernard Lyot Telescope (the principal investigators: KU and Claude Catala), for 
which the data are now public\footnote{\url{http://www.cfht.hawaii.edu/ObsInfo/Archive},\\ \hspace*{8pt}\url{http://tblegacy.bagn.obs-mip.fr/narval.html}}. Those two instruments observed solar-type 
stars with the widest range of brightness: $12>V>7$~mag. 

The total number of spectra which we analyse is~189. However, because 15 stars were observed with two instruments and one star, with three, the number of the 
individual stars that we discuss in this paper is~172. Three of those stars are double-lined spectroscopic binaries (SB2) and therefore, we do not compute their 
atmospheric parameters; those values are provided for 169 stars. The stars with multiple observations are used for an internal check of the consistency of our results. 
Those for which $T_{\rm eff}$,  $\log g $, and [Fe/H] have been derived by \citet{bruntt2012} or \citet{thygesen2012} from the ESPaDOnS and NARVAL spectra are 
included for the sake of analysing possible differences in the results obtained by means of different methods.

124 stars from our sample have been recently discovered to show solar-like oscillations and eleven, to show other types of photometric variability (see Table~\ref{atmos} 
and the references therein.) Four stars fall into the field of the open cluster NGC~6811 (KIC~9655101, 9655167, 9716090, and 9716522) and three, into the field 
of NGC~6819 (KIC~5024851, 5112786, and 5199859.) KIC~3632418 (=~Kepler~21b) is a planet-hosting star \citep{howell2012} while KIC~8866102, 9414417, 9955598, and 
10963065 are {\it Kepler} candidates for stars with planets\footnote{\url{http://planetquest.jpl.nasa.gov/kepler/table}}. 

\section{Observations}
\label{observations}

Our programme stars were observed with five different instruments. In Table~\ref{instruments}, we provide names of those instruments, the names of the 
telescopes, the acronyms of the observatories, the number of acquired spectra ($n$), the year in which the data were acquired, the spectral range and the resolving 
power ($R$) of the spectrograph, the exposure time, and the typical signal-to-noise ratio ($S/N$) along with the location in the wavelength where it was measured. 

For all the instruments, the bias, flat field, and calibration lamp measurements were acquired in the evening and the morning. For FIES, additional spectra of the calibration 
lamp were obtained before each science observation. The data were reduced and calibrated following standard reduction procedures which included subtraction of the bias frame, 
correction for flat field, extraction of the orders, wavelength calibration, and cleaning the spectrum from cosmic rays. The normalisation 
of the spectra to the level of unity was done manually with {\sf IRAF}\footnote{IRAF is distributed by the National Optical Astronomy Observatory, which is 
operated by the Association of Universities for Research in Astronomy, Inc.}. More details about observations carried out with each of the five instruments are 
provided below.

\begin{table*}
\begin{center}
\caption{A summary of instruments and observations.}
\label{instruments}
{\small
\begin{tabular}{lllrclcrrc}
\hline\hline\noalign{\smallskip}
Instrument & Telescope & Observatory & n  & Year of     & Spectral      & R      & $t_{\rm exp}$~~~~ & $S/N$~~~~~~~~    \\
           &           &             &    & observations & range [{\AA}] &        & [s]~~~~~          &                  \\
\hline                                   
\hline                                    
FIES       & NOT       & ORM         &  4 & 2010-2011   & 3700-7300     & 46,000 &  420-2050         & 100 at 4900 \AA  \\
FIES       & NOT       & ORM         &  6 & 2010-2011   & 3700-7300     & 25,000 & 1500-2600         & 100 at 4900 \AA  \\
FRESCO     & 91-cm     & INAF-OACt   & 18 & 2009-2010   & 4300-6800     & 21,000 & 2700-4200         &  80 at 6500 \AA  \\
HERMES     & Mercator  & ORM         & 20 & 2010-2011   & 3800-9000     & 85,000 &  500-2600         &  90 at 6500 \AA  \\
NARVAL     & TBL       & OPM         & 50 & 2010        & 3700-10500    & 75,000 &  $<900$           & 100 at 5200 \AA  \\
ESPaDOnS   & CFHT      & MKO         & 91 & 2010        & 3700-10500    & 80,000 &  $<900$           & 100 at 5200 \AA  \\
\hline
\end{tabular}
}
\end{center}
\end{table*}

\subsection{FIES}
FIES (FIber-fed Echelle Spectrograph) is a cross-dispersed high-resolution echelle spectrograph mounted on the 2.56-m Nordic Optical Telescope (NOT) at the
Observatorio Roque de los Muchachos (ORM) on La Palma, Spain. We used the medium-resolution mode ($R=46,000$) to observe the bright stars ($10>V>7$~mag), and the 
low-resolution mode ($R=25,000$), for the faint ones ($11.5>V>10$~mag). The observations were carried out by EN and JL. The spectra were 
reduced and calibrated using the dedicated reduction software {\it FIEStool} \citep{stempels2004} that is based on existing standard {\sf IRAF} reduction procedures.

\subsection{FRESCO}
FRESCO is a fiber-linked REOSC echelle spectrograph fed by the 91-cm telescope of the Osservatorio Astrofisico di Catania (INAF-OACt), Italy. The 
observations were carried out by JM-\.Z. The data were reduced and calibrated with {\sf IRAF}.

\subsection{HERMES}
HERMES is a fiber-fed echelle spectrograph attached to the Flemish 1.2-m telescope Mercator, also at the ORM (Spain). It is optimised for high resolution, 
stability, and broad wavelength coverage which is achieved primarily by implementing an image slicer, an anti-fringe CCD coating, and a thermal enclosure 
\citep{Raskin2011}. The observations were carried out by DD, PL, JG, NG, DV, SB, and CJ. The 
data reduction and calibration were performed with a dedicated Python-based pipeline \citep{Raskin2011}. 

\subsection{ESPaDOnS and NARVAL}
The ESPaDOnS and the NARVAL spectrographs are very similar to each other. ESPaDOnS is mounted at the 3.6-m Canada-France-Hawaii Telescope (CFHT) at Mauna Kea 
Observatories (MKO, USA) while NARVAL is mounted at the 2-m T\'elescope Bernard Lyot (TBL) at the Observatoire Pic du Midi (OPM, France). Both instruments 
observed the {\it Kepler} targets in service mode. Data used in the present paper are available in the public archive of the Canada-France-Hawaii Telescope (CFHT) 
Science Data Archive and the CNRS/INSU CDAB/Bass2000 TBLegacy database. They were reduced and calibrated as part of the service programme by means of the data 
reduction software Libre-ESpRIT written and provided by J.-F.\ Donati from IRAP, Observatoire Midi-Pyr\'en\'ees \citep{Donati1997}. 

\section{Methods of analysis}
\label{tgmet}
We use two different methods of the spectroscopic analysis, {\sf ROTFIT} and {\sf ARES+MOOG}, to derive the atmospheric parameters for our programme stars. As
described below, each of these methods makes use of a different approach and has different limitations. 

\subsection{ROTFIT}

\begin{figure}
\includegraphics[width=8.5cm,angle=0]{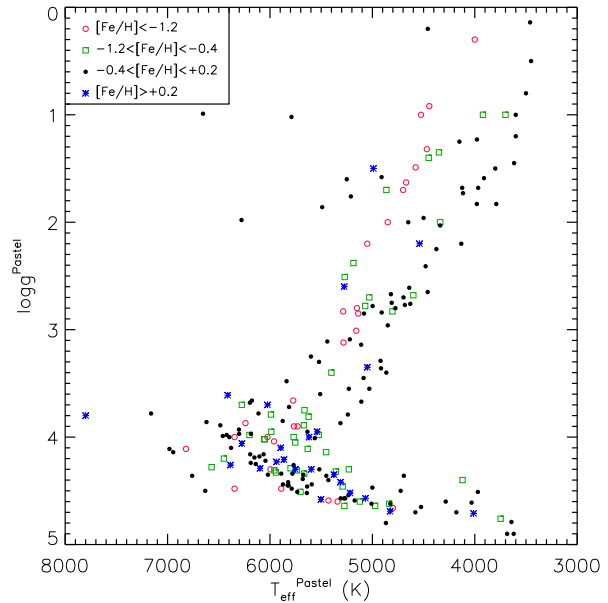}
\caption[]{Distribution of the parameters adopted for the reference stars in a $T_{\rm eff}$--$\log g$ plane.
Stars in different ranges of metallicity are displayed with different colours and symbols.}
\label{fig:hr_stand}
\end{figure}

\begin{figure*}
\includegraphics[width=5.7cm,angle=0]{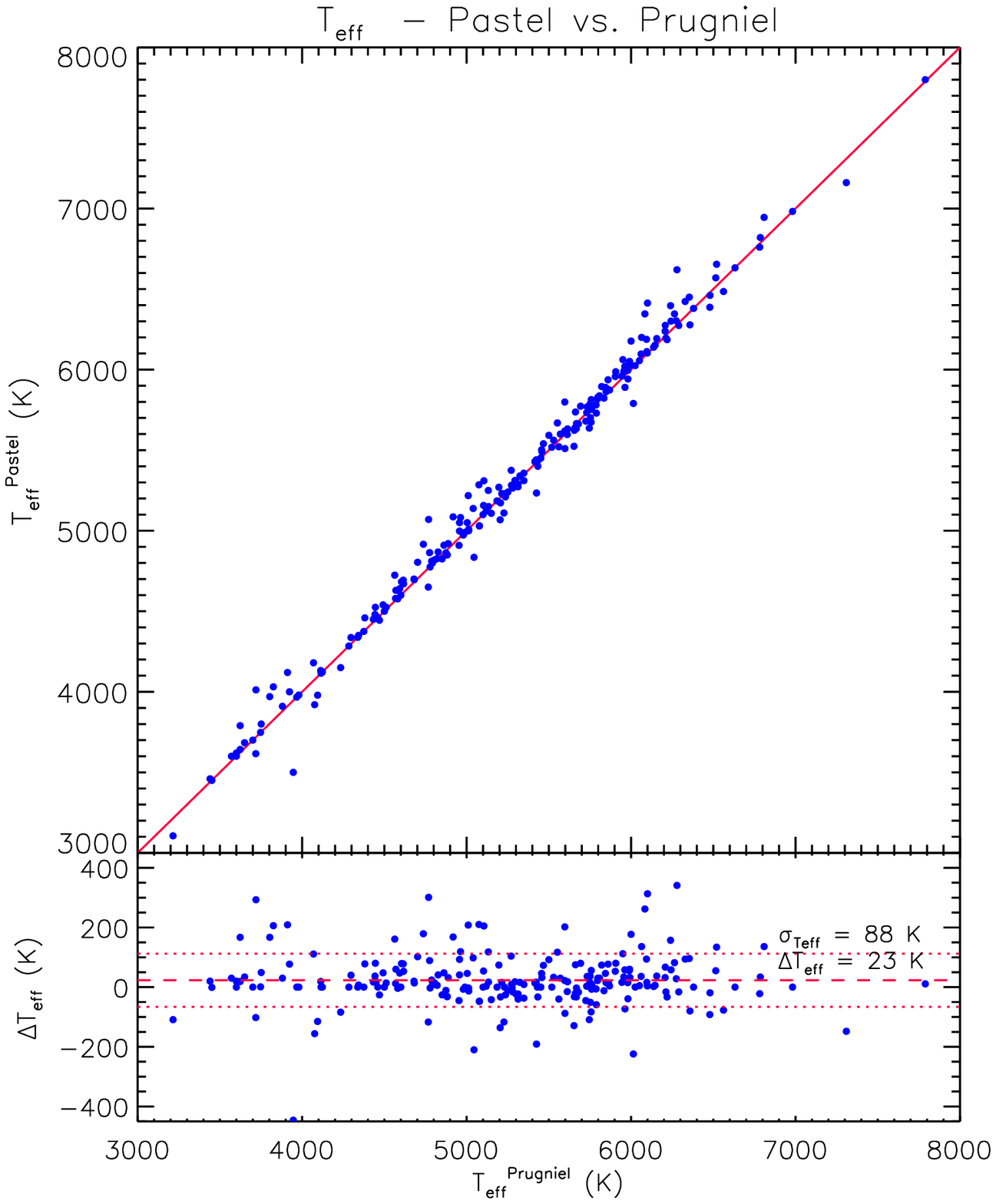}
\includegraphics[width=5.7cm,angle=0]{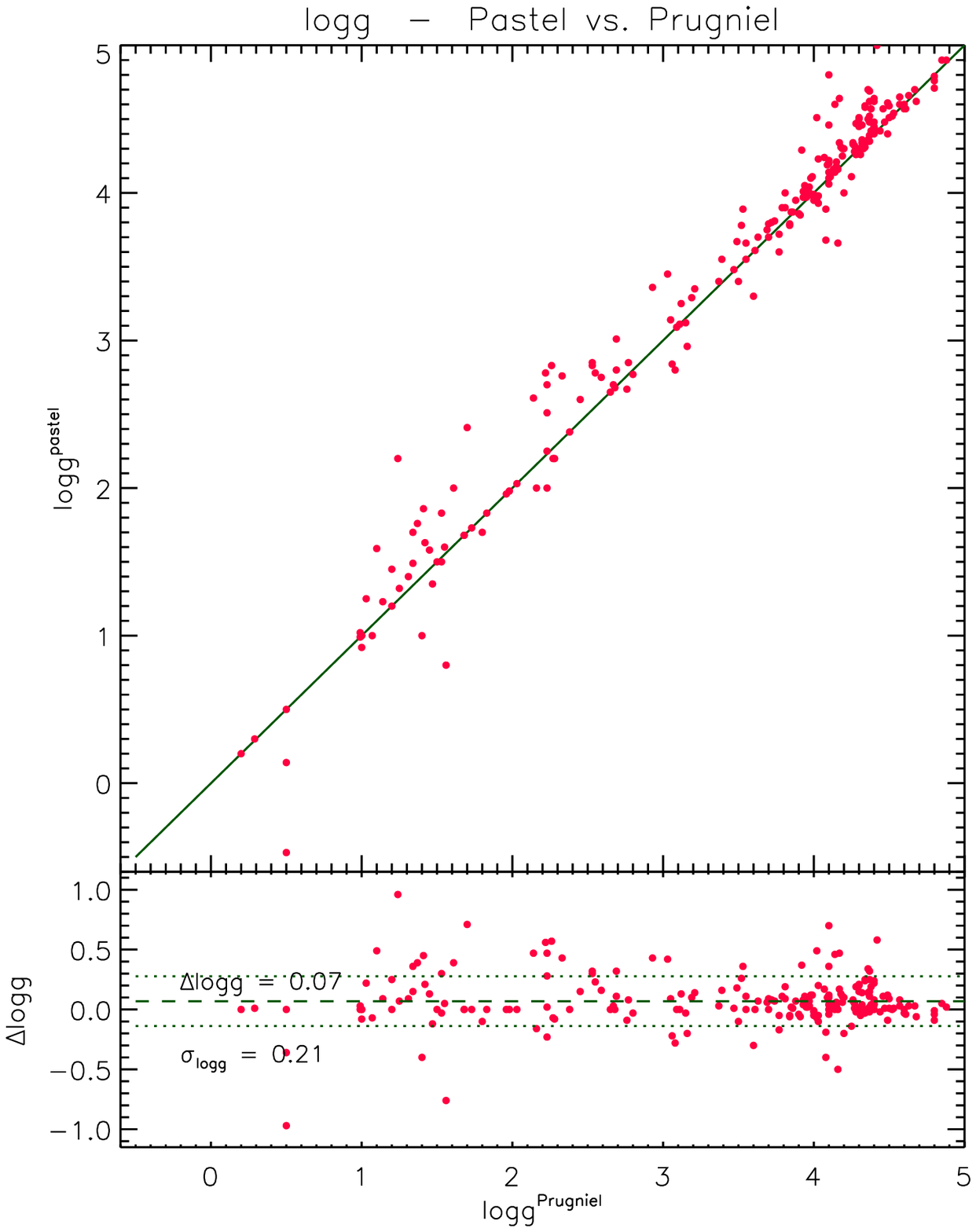}
\includegraphics[width=5.7cm,angle=0]{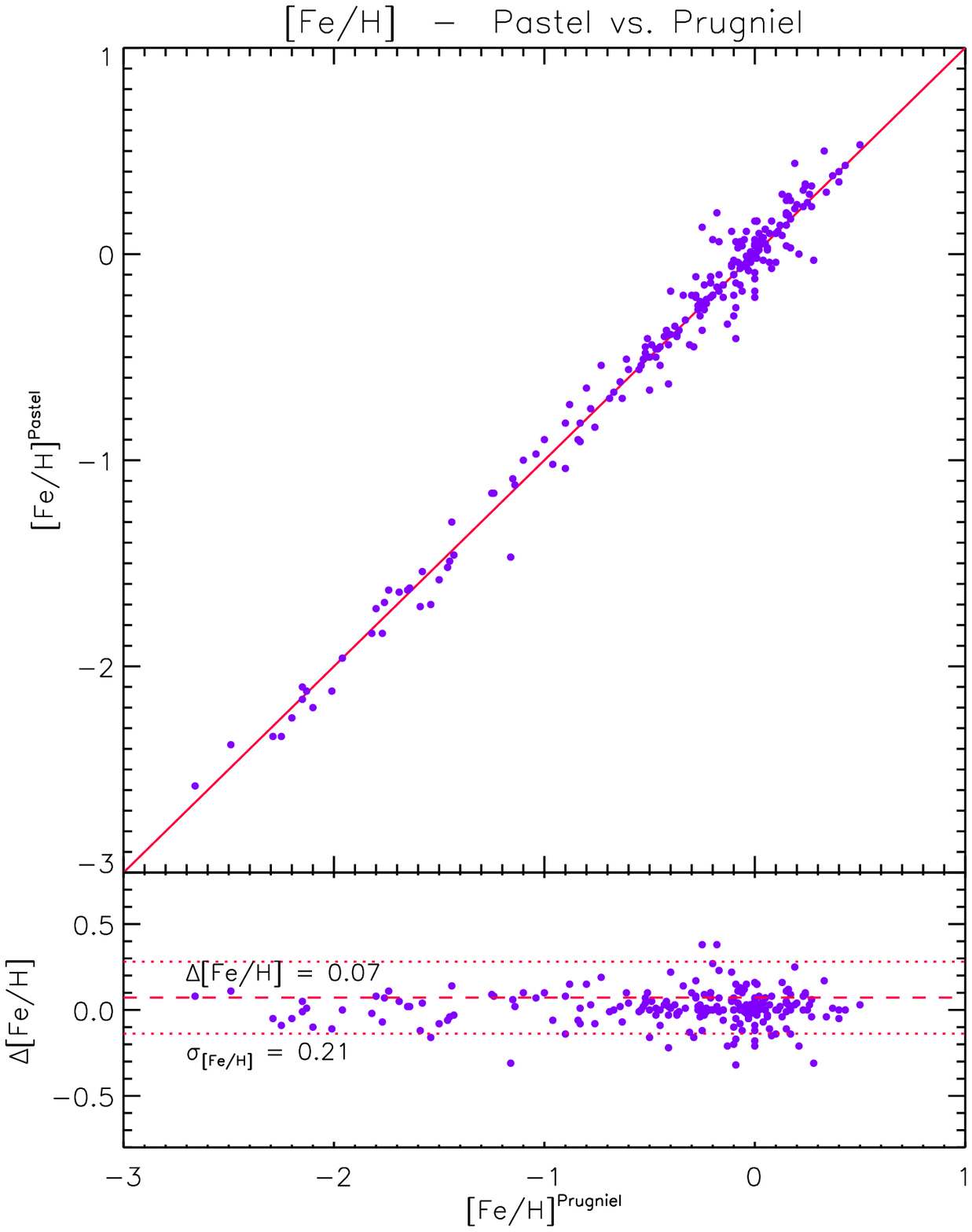}
\caption[]{Comparison between the adopted atmospheric parameters (from the PASTEL catalogue, \citealt{Soubiran2010}) and those from
the ELODIE library v3.1 \citep{Prugniel2007}.}
\label{fig:comp_prugn}
\end{figure*}

\begin{figure}
\includegraphics[width=8.5cm,angle=0]{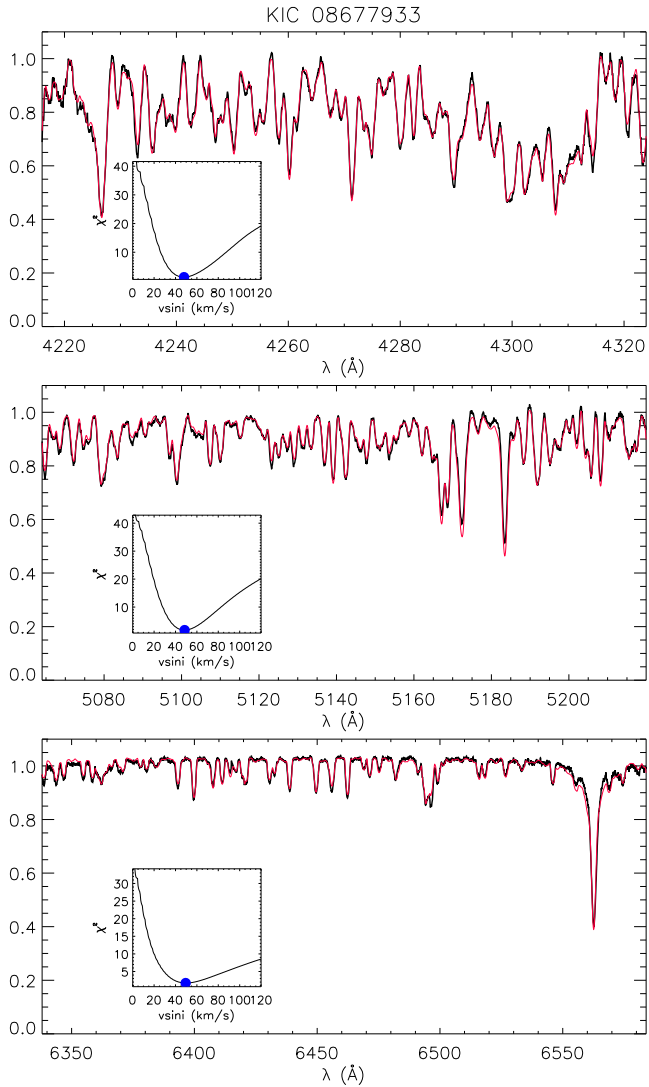}
\caption[]{A part of the output of the {\sf ROTFIT} code for  KIC~8677933, a G0IV fast rotating star ($v\sin i = 49.6$~km\,s$^{-1}$) 
in three different spectral segments. In each panel,  the best template spectrum (thin red line), rotationally broadened and Doppler-shifted, is superimposed 
onto the observed spectrum of KIC~8677933 (thick black line). The insets show the $\chi^2$ for the best template as a function of $v\sin i$.}
\label{fig:vsini_chi}
\end{figure}
The code {\sf ROTFIT}, developed by \citet{Frasca2003, Frasca2006} in IDL\footnote{IDL (Interactive Data Language) is a registered trademark of ITT 
Visual Information Solutions.}  software environment, was originally designed to perform an automatic MK spectral classification and a $v \sin i$ measure by 
comparing the target spectrum with a grid of slowly-rotating reference spectra. The latter are resampled, aligned to the target spectrum by cross-correlation, 
and rotationally broadened by convolution with a rotational profile of increasing $v \sin i$ until the 
minimum of $\chi^2$ is reached. The $\chi^2$ is defined by the following equation:
\begin{equation}
\chi^2 = \frac{1}{N}\sum_j \frac{(y_j^{\rm obs}-y_j^{\rm templ})^2}{\sigma_j^2}
\end{equation}
\noindent{where $y_j^{\rm obs}$ is the value of the continuum-normalised flux of the observed spectrum at the $j^{\rm th}$ point, $y_j^{\rm templ}$
is the corresponding value for the rotationally-broadened template, $\sigma_j$ is the error of $y_j^{obs}$, and $N$ the number of data points.}

The code was subsequently used for evaluating, in addition to $v \sin i$, the atmospheric parameters $T_{\rm eff}$, $\log g$, and $\rm [Fe/H]$ by 
adopting a list of reference stars with well known parameters \citep[see, e.g.,][]{molenda2007,guillout2009}. A good agreement between the values of $T_{\rm eff}$, 
$\log g$, and $\rm [Fe/H]$ derived with {\sf ROTFIT} and those obtained with other techniques are shown, e.g., by \citet{metcalfe2010,frohlich2012}.   
Unlike codes based on the measurements of equivalent widths and curves of growth, {\sf ROTFIT} can also be applied to the spectra of FGK stars with 
$v \sin i$ exceeding 20 km\,s$^{-1}$ or spectra with a moderate resolution, for which the blending of individual lines prevents the use of the previous methods. 
Our tests done on a selected sample of spectra of slowly rotating stars artificially broadened by convolution with a rotation profile showed that the derived 
atmospheric parameters are not significantly affected up to $v\sin i\approx50$\,km\,s$^{-1}$. For higher values of $v\sin i$, the uncertainties of the atmospheric 
parameters increase and eventually become as high as 1.5--2 times the original values for $v\sin i=100$\,km\,s$^{-1}$. As an example of the output produced by 
{\sf ROTFIT} for a fast-rotating star (KIC~8677933, G0IV, $v\sin i = 49.6$~km\,s$^{-1}$), the plot of the observed spectrum and the best-matching template in 
three spectral regions is shown in Fig.~\ref{fig:vsini_chi}. The insets, in which $\chi ^2$ for the best template is plotted as a function of $v\sin i$, show that 
even for such a fast-rotating star the minimum of $\chi ^2$ is well defined and that the observed and the template spectra match well.

Our library of reference stars consists of 221 high-resolution ($R$ = 42,000), high-$S/N$ spectra of slowly rotating stars acquired with the fiber-fed 
echelle spectrograph ELODIE at the Haute-Provence Observatory which are available in the ELODIE archive \citep{Prugniel2001}. The atmospheric parameters
of these stars were retrieved from the PASTEL catalogue \citep{Soubiran2010} and are listed in Table~\ref{rot-lib}.
Since for most stars in the PASTEL catalogue there are several determinations of $T_{\rm eff}$, $\log g$, and $\rm [Fe/H]$ obtained by different authors and with 
different methods, we selected those which are recent, based either on spectral synthesis or on the analysis of the equivalent widths, and which are not significantly 
different from the bulk of the other determinations. The $T_{\rm eff}$--$\log g$ diagram (Fig.~\ref{fig:hr_stand}) shows that the reference stars are rather well distributed 
in all the regions relevant for FGK stars with a density that is not certainly uniform, but not very far from it. A comparison between the values of $T_{\rm eff}$, 
$\log g$, and $\rm [Fe/H]$ adopted by us and those homogeneously re-determined in the ELODIE library~v3.1 by \citet{Prugniel2007} for 220 stars (we note that 
the latter authors do not provide atmospheric parameters for star GJ~166~C) displays very small offsets and dispersions of 88~K, 0.21~dex, and 0.21~dex for $T_{\rm eff}$, 
$\log g$, and $\rm [Fe/H]$, respectively (see Fig.~\ref{fig:comp_prugn}). That ensures that the adopted parameters are well consistent with those of \citet{Prugniel2007}.
Table~\ref{rot-lib} gives also the MK types of our reference stars. Those were adopted either from the SIMBAD database or from the General Catalogue of Stellar Spectral 
Classifications by \citet{skiff2013}. For most stars that classification agrees with the MK types given for our reference stars in the ELODIE database.

\begin{table*}
\begin{center}
\caption{The first three rows and the last row of Table~\ref{rot-lib} which provides the MK type, effective temperature, surface gravity, and metallicity of 
221~reference stars used by the {\sf ROTFIT} code. The atmospheric parameters are adopted mostly from the PASTEL catalogue \citep{Soubiran2010}. The last column provides  
the references to the sources of the adopted values. The full table is available only in an electronic form.}
\label{rot-lib}
{\small
\begin{tabular}{rllrrrl}
\hline\hline\noalign{\smallskip}
No   & Star Name  & Spectral Type & $T_{\rm eff}$ & $\log g$ & $\rm [Fe/H]$  & Reference to the Literature\\
\hline                                   
\hline                                    
   1 &  BD+023375 & F9IVsub &     5960 &     4.04 &    $-$2.34 &  Stephens \& Boesgaard (2002), AJ, 123, 1647\\
   2 &  BD+044551 & F7Vw    &     5730 &     3.90 &    $-$1.70 &  Tomkin et al. (1992), AJ, 104, 1568\\
   3 &  BD+174708 & sdF8    &     6025 &     4.00 &    $-$1.63 &  Fulbright (2000), AJ, 120, 1841\\
\multicolumn{7}{c}{\dotfill}\\                                                                
 221 &  HD345957  & G0Vwe   &     5766 &     3.90 &    $-$1.46 &  Gratton et al. (2003), A\&A, 404, 187\\

\hline
\end{tabular}
}
\end{center}
\end{table*}

We have analysed independently spectral segments of 100~{\AA} each or the individual echelle orders, depending on the format of the spectra. We have 
excluded from the fit the spectral regions heavily affected by telluric lines, like the $\rm O_2$ band from 6275 to 6330~{\AA}. Per each segment, we took the 
average parameters of the best ten reference stars ($\sim$\,5\,\% of the total sample), with a weight proportional to $\chi^{-2}$. Although {\sf ROTFIT} uses a 
fixed number of nearest neighbours (10), the weight provided by the $\chi^{2}$ limits the contamination of the final parameters and allows a meaningful estimate 
of the uncertainties.
 
We adopted as the best estimates of $T_{\rm eff}$, $\log g$, and $\rm [Fe/H]$ the weighted averages of the results of each spectral segment using 
$\sigma_i^{-2}\chi_i^{-2} f_i$ as the weight. Here, $\sigma_i$ is the standard error of the parameters for the best 10 templates of the $i^{\rm th}$ spectral 
segment. As such it is adopted as a relative measure of the internal consistency for the $i^{\rm th}$ spectral segment/order, $\chi_i^{-2}$ is the minimum 
chi-square of the $i^{\rm th}$ segment and takes into account differences between orders due to the $S/N$ and the goodness of the fit. The factor $f_i$, which 
is an integral over all the $i^{\rm th}$ spectral segment of $(F_{\lambda}/F_{\rm C}-1)$, is proportional to the total line absorption and was included
in the weight to correct for the different amount of information contained in different spectral segments. We evaluated the uncertainties of $T_{\rm eff}$, 
$\log g$, $\rm[Fe/H]$, and $v \sin i$ as the standard errors on the weighted means to which we have summed in quadrature the average dispersion of differences between 
the stellar parameters of our reference stars given in the PASTEL catalogue and in \citet{Prugniel2007}, i.e., $\sigma_{T_{\rm eff}}= 88$~K, $\sigma_{\log g}=0.21$\,dex, and 
$\sigma_{\rm [Fe/H]}=0.21$\,dex (see Fig.~\ref{fig:comp_prugn}). The MK classification of the target star is performed by 
adopting the spectral type and the luminosity class of the reference star which more frequently matched with the target spectrum in the different spectral segments.

\subsection{ARES+MOOG}
\label{aresmoog}

This method of analysis allows the derivation of $T_{\rm eff}$, $\log g$, the microturbulence $\xi_{\rm t}$, and $\rm [Fe/H]$ following a procedure described and 
used in \citep{santos2004, sousa2006, sousa2008, sousa2011a, sousa2011b}. Because this method relies on two core codes, namely {\sf ARES (Automatic Routine for 
line Equivalent widths in stellar Spectra)} developed by \citet{sousa2007} and {\sf MOOG} developed by \citet{sneden1973}, we refer to it as to {\sf ARES+MOOG}. 
The method is based on measuring equivalent widths ($EW$s) of Fe~I and Fe~II weak absorption lines and then imposing excitation and ionisation equilibrium, 
assuming LTE approximation. The 2002 version of the code {\sf MOOG} code is used together with the grid of Atlas 9 plane-parallel model atmospheres \citep{kurucz1993}. 
In this procedure, [Fe/H] is a proxy of the metallicity. The equivalent widths are measured automatically with the {\sf ARES} code which successfully reproduces 
the manual, interactive determination of $EW$s.

One of the unique characteristics of {\sf ARES+MOOG} is the list of iron lines. Although a preliminary large list of nearly 500 lines was compiled from the Vienna 
Atomic Line Database {\citep{kupka1999}, the final list includes around 300 most reliable lines that were carefully tested when automatically measured with {\sf 
ARES} \citep{sousa2008}. Another important aspect of the list is the adopted atomic parameters for each line: The oscillator strengths ($\log gf$) of the lines 
were recomputed through an inverse analysis of the solar spectrum, allowing in this way to perform a differential analysis relatively to the Sun.

The errors of the parameters derived with {\sf ARES+MOOG} were obtained by quadratically adding 60~K, 0.1 and 0.04~dex to the method's intrinsic errors in 
$T_{\rm eff}$, $\log g$, and [Fe/H], respectively. Those values were obtained by measuring the typical standard deviation of the parameters discussed by 
\citet{sousa2008}. A~more complete discussion of errors representative for this spectroscopic method can be found in \citet{sousa2011a}.

Since we adopt a differential analysis (using the Sun as the reference), this method is expected to work very well for solar-type stars and to be less accurate 
for the cooler and the hotter stars, and those which are significantly different from the Sun. For this reason, we don't provide results for stars cooler than 
about 4,500~K. Moreover, since {\sf ARES+MOOG} requires precise measurements of the $EW$s, we do not provide results for stars with $v\sin i > \rm 
10~km\,s^{-1}$ because higher rotation causes line blending, preventing precise determination of $EW$. Finally, as {\sf ARES+MOOG} works best with high-resolution 
spectra, we do not apply this method to stars observed with $R \le 25,000$.

\section{Atmospheric parameters}
\label{results}
The values of $T_{\rm eff}$~[K], $\log g$~[cm\,s$^{-2}$], $\rm [Fe/H]$~[dex], and $v\sin i$~[km\,s$^{-1}$] with their standard deviations, and the MK type derived 
with {\sf ROTFIT} are listed in columns 2-10 of Table~\ref{atmos}. The values of $T_{\rm eff}$~[K], $\log g$~[cm\,s$^{-2}$], $\rm [Fe/H]$~[dex], and 
$\xi_{\rm t}$~[km\,s$^{-1}$] with their standard deviations derived with {\sf ARES+MOOG} are listed in columns 11-18. KIC numbers are provided in the first column 
and the designations of the instrument, in the last but one. The last column contains information about the type of variability of the stars. We use boldface 
font for KIC numbers of the stars for which atmospheric parameters are derived for the first time. The boldface font instrument designations indicate that the 
respective spectrum has not been used in any previous analyses in the literature.

For KIC~9693187 and 11179629, we detected lines of both components in the spectrum. KIC~9025370 was discovered to be a double-lined spectroscopic binary by  
Thygesen et al. (in preparation). We do not compute the atmospheric parameters for these three stars and we indicate in Table~\ref{atmos} that they are SB2 systems.

For KIC~6370489, 10709834, and 10923629 we do not provide the atmospheric parameters obtained with {\sf ARES+MOOG}. In the spectrum of the first star we find 
too few useful spectral lines for {\sf ARES+MOOG} to converge. In case of KIC~10709834 and 10923629, {\sf ARES+MOOG} yields very high values of $\log g$ which are not 
confirmed with {\sf ROTFIT}. Therefore, we suspect that the results produced by {\sf ARES+MOOG} for those two stars may be spurious.

Below, we discuss the values of $T_{\rm eff}$, $\log g$, and [Fe/H] determined with {\sf ARES+MOOG} and with {\sf ROTFIT}. We compare these results with each other 
and with those obtained with the {\sf VWA (Versatile Wavelength Analysis)} code by \citet{bruntt2012} and \citet{thygesen2012} for one hundred stars from our sample. 
For 145 stars, we compare our temperature determinations with the ones derived with the infrared flux method (IRFM) by \citet{pinsonneault2012}.

\subsection{Effective temperature}
As shown in Fig.~\ref{allteff}, the differences between the values of $T_{\rm eff}$ derived with {\sf ARES+MOOG}, {\sf ROTFIT}, {\sf VWA} and {\sf IRFM} show various
offsets and large standard deviation. The standard deviation is lowest but still significant when the comparisons involve $T_{\rm eff}$ computed with {\sf VWA} 
(Fig.~\ref{allteff}~$b$, $d$, and~$f$). This must be related to the fact that {\sf VWA} was applied to high-$S/N$, high-resolution spectra from ESPaDOnS and NARVAL: 
When data of high quality are used, all methods yield $T_{\rm eff}$ which are more precise and accurate

For stars with $T_{\rm eff}>6,000$~K, the effective temperatures derived with {\sf ARES+MOOG} are systematically hotter than those obtained with {\sf 
ROTFIT} and {\sf VWA} (Fig.~\ref{allteff}~$a$ and~$b$.) Between 5,000~K and 6,000~K these three methods agree well but for stars cooler than 5,000~K, 
{\sf ARES+MOOG} yields slightly higher values of $T_{\rm eff}$ which is why for the coolest stars the agreement between {\sf ARES+MOOG} and {\sf ROTFIT} or {\sf 
VWA} is worse again. The reason for this may be related to the selection of spectral lines. The original list of lines is optimised for solar-type stars while 
for cool stars many of those lines are affected by blending. This effect contributes strongly to the observed offset in temperature. A refinement of the selection 
of lines to produce consistent results for stars cooler than 5,000~K will be presented by \citet{tsantaki2013}.

Fig.~\ref{allteff} $a$ and $b$ show that when {\sf ROTFIT} and {\sf VWA} are compared to {\sf ARES+MOOG}, the differences show a similar pattern. This 
suggest that $T_{\rm eff}$ obtained with {\sf ROTFIT} and {\sf VWA} should be close to each other. Indeed, the mean difference between $T_{\rm eff}$ derived by 
means of those two methods is relatively small, only 70~K. Nevertheless, the standard deviation of the differences, 123~K, is still high (Fig.~\ref{allteff}~$d$.)

\begin{figure*}
\includegraphics[width=17cm,angle=0]{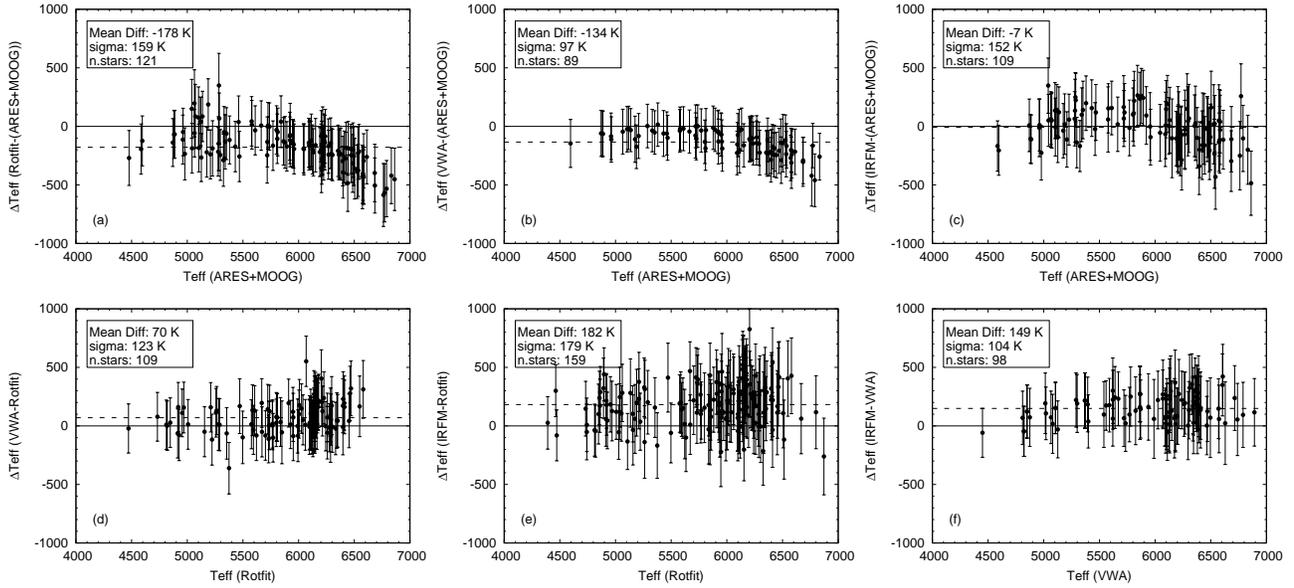}
\caption[]{Mutual comparison of $T_{\rm eff}$ values measured with four different methods: {\sf ROTFIT} and {\sf ARES+MOOG} (this paper), {\sf VWA} 
\citep{bruntt2012, thygesen2012} and {\sf IRFM} \citep{pinsonneault2012}. In the insets, we give the mean difference between the compared sets of data, the 
standard deviation of the mean, and the number of stars in common.} 
\label{allteff}
\end{figure*}

\begin{figure*}
\includegraphics[width=17cm,angle=0]{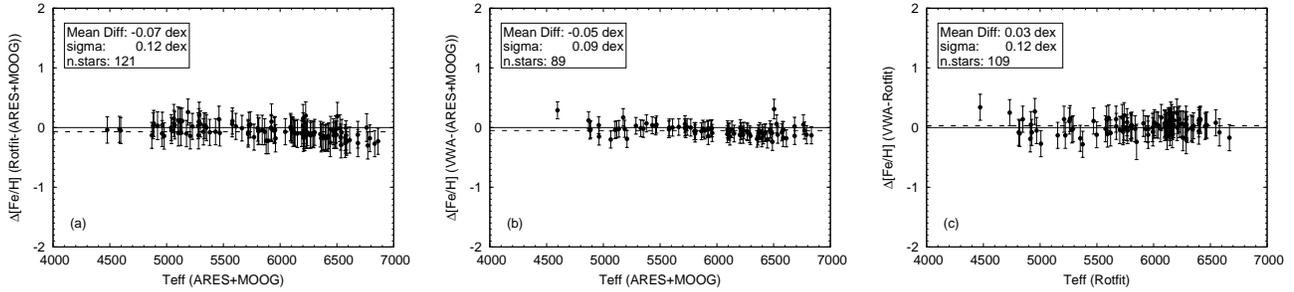}
\caption[]{Mutual comparison of [Fe/H] values measured with three different methods: {\sf ROTFIT} and {\sf ARES+MOOG} (this paper), and {\sf VWA} \citep{bruntt2012,
thygesen2012}. In the insets, we give the mean difference between the compared sets of data, the standard deviation of the mean, and the number of stars in common.}
\label{feh}
\end{figure*}

\begin{figure*}
\includegraphics[width=17cm,angle=0]{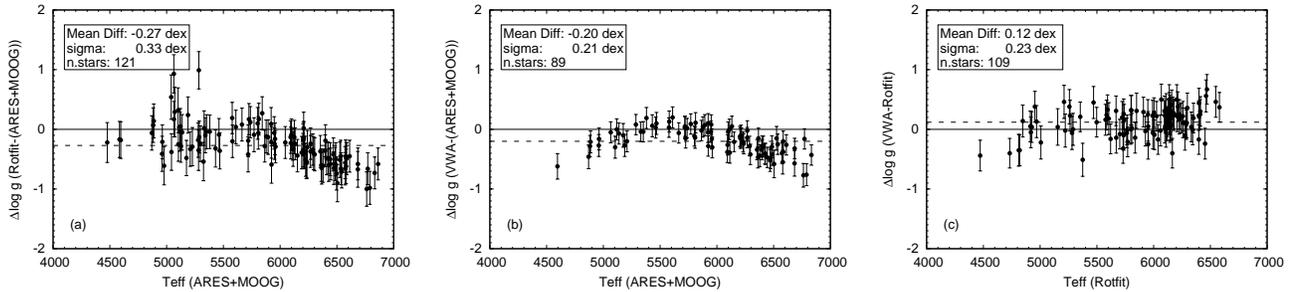}
\caption[]{The same as in Fig.~\ref{feh} but for the $\log g$ values.}
\label{logg}
\end{figure*}

When compared with the {\sf IRFM}-based $T_{\rm eff}$ measured by \citet{pinsonneault2012}, the values of $T_{\rm eff}$ derived with {\sf ARES+MOOG} show a 
negligible offset of 7~K but a high standard deviation of 152~K (Fig.~\ref{allteff} $c$). The two other methods, {\sf ROTFIT} and {\sf VWA}, show a much higher 
mean difference, 182 and 149~K, and similar standard deviation of 179 and 104~K, respectively (Fig.~\ref{allteff} $e$ and $f$). Therefore, it is difficult to 
say which of those methods, if any, agrees with {\sf IRFM} best.

Since {\sf ARES+MOOG} temperature scale is known to be in a very good agreement with {\sf IRFM} \citep[see][]{sousa2008}, we would expect the results shown in 
Fig.~\ref{allteff}~$c$ to agree much better than is the case. One of the plausible explanations of the observed scatter is the fact that the {\sf IRFM}-based 
$T_{\rm eff}$ provided by \citet{pinsonneault2012} were derived only from one colour index, $(J-K_S)$. This index produces the values of $T_{\rm eff}$ which
show the highest scatter when compared with the {\sf IRFM} $T_{\rm eff}$ derived from other colour indices \citep[see Figs.~9-11 in][]{pinsonneault2012}. 
Indeed, when the mean values of the {\sf IRFM} $T_{\rm eff}$ of stars in the Hyades and 
Praesepe open clusters (computed from $(B-V)$, $(V-I_C)$, and $(V-K_S)$ indices) are compared with the $(J-K_S)$-based $T_{\rm eff}$ values, and their differences 
are plotted as a function of the YREC $T_{\rm eff}$ computed by \citet{pinsonneault2012} by using the isochrones by \cite{an2007}, the differences are positive by 
around 100~K for stars which are cooler than 5900~K and negative by around 50~K for stars which are hotter than 6200~K \citep[see the 
Fig.~13 in][]{pinsonneault2012}. The $(J-K_S)$-based $T_{\rm eff}$ of the {\it Kepler} stars shown in the same figure show a similar trend as the $(J-K_S)$-
based $T_{\rm eff}$ of stars in Hyades and Praesepe; only for stars hotter than 6200~K their values are closer to the YREC $T_{\rm eff}$ scale (but still lower 
by around 100~K.) 

These results allow us to conclude that the high standard deviation of the differences shown in Fig.~\ref{allteff}~$c$ is likely due to the calibration 
issues in the $(J-K_S)$ colour index \citep[c.f. Sect. 3.3 in][]{pinsonneault2012}, not to any possible weakness of {\sf ARES+MOOG}. The general consistency of 
the effective temperatures derived from spectroscopy and the {\sf IRFM} method is supported also by \citet{bruntt2012} who show that $T_{\rm eff}$ computed with 
{\sf VWA} are in a good agreement with the {\sf IRFM} $T_{\rm eff}$ values derived from the $V_T - K_S$ index and the calibration of \citet{casagrande2010} as 
the mean difference between those two scales of temperature is only $4\pm85$~K \citep[see][figure 2]{bruntt2012} whereas the standard deviation of the differences 
between {\sf VWA}-based $T_{\rm eff}$ and those obtained from the $(J-K_S)$ index shown in Fig.~\ref{allteff}~$f$ is comparable to the value obtained for
Fig.~\ref{allteff}~$c$.

One should also keep in mind that {\sf IRFM} $T_{\rm eff}$ derived by \citet{pinsonneault2012} may be slightly affected by interstellar reddening of the 
stars: Because there are no individual measurements of $E(B-V)$ for each target, \citet{pinsonneault2012} correct the observed magnitudes for interstellar 
extinction using the map-based estimates of extinction from KIC. Unfortunately, those values of $E(B-V)$ are not accurate as has been shown by \citet{molenda2009} 
for 29 nearby ($16<r<240$~pc), bright ($9.0<V<11.2$) {\it Kepler} targets which were observed photometrically by those authors. \citet{molenda2009} did not find 
any evidence that those stars were reddened while $E(B-V)$ provided in KIC were sometimes as high as 0.06~mag. The influence of inaccurate $E(B-V)$ used by 
\citet{pinsonneault2012} on the {\sf IRFM} $T_{\rm eff}$ may be small but should be considered as one of possible sources of the scatter in Fig.~\ref{allteff}~$c$, 
$e$, and $f$.

\subsection{Metallicity}
As shown in Fig.~\ref{feh} $a$, $b$, and $c$, the values of [Fe/H] derived with {\sf ARES+MOOG}, {\sf ROTFIT} and {\sf VWA} agree with each other to within the error 
bars for almost all targets. The mean differences between these determinations do not exceed 0.07~dex. However, their standard deviations are quite large and 
comparable to the typical uncertainty of [Fe/H] derived with {\sf ROTFIT} or to twice the typical uncertainty of [Fe/H] derived with {\sf ARES+MOOG}. 

For the stars hotter than 6,000~K, the values of [Fe/H] derived with {\sf ARES+MOOG} are slightly higher than those obtained with {\sf ROTFIT} or {\sf VWA} 
(Fig.~\ref{feh}~$a$ and $b$). However, this trend does not affect the overall consistency of the results. The values of [Fe/H] obtained with {\sf ROTFIT} and {\sf VWA} 
agree best (Fig.~\ref{feh} $c$) showing a mean difference of 0.03~dex and no trends at high temperatures. The high standard deviation is not reduced, however, and 
it is as high as that in Fig.~\ref{feh} $a$ where the mean difference is the highest and the trend at the high temperatures is most obvious.

\subsection{Surface gravity}
The surface gravity is the parameter which is least constrained when derived with {\sf ARES+MOOG}. The reason for this is related to the number of iron lines used
in the method. Although we use nearly 300 Fe~I lines, which constrains very well the temperature, microturbulence, and the metal abundance, $\log g$, which comes 
from the ionisation balance, requires an analysis of Fe~II lines. Unfortunately, the number of Fe~II lines in our line list is limited to less than 20. Due to 
that small number, the results of the analysis are more sensitive to errors and more uncertain.

The differences between the $\log g$ values computed with {\sf ARES+MOOG}, {\sf ROTFIT}, and {\sf VWA} illustrated in Fig.~\ref{logg} $a$ and $b$, are around 0.2~dex, and 
show discrepancies increasing for hot stars. The trends visible in Fig.~\ref{logg} $a$ and $b$, mimic those in Fig.~\ref{allteff}~$a$ and $b$ which may be a result of 
strong correlations between $T_{\rm eff}$ and $\log g$. The values of $\log g$ obtained with {\sf ROTFIT} and with {\sf VWA} agree with each other better (Fig.~\ref{logg}~$c$.) 
The mean difference between them is the lowest, 0.12~dex, and there are no trends for hot stars. However, the standard deviation of these differences is still high.

\section{Discussion}
\label{discussion}
Our analysis shows that deriving precise and accurate atmospheric parameters is not a trivial task and that limiting such determinations to one method can result in a
false impression that the accuracy of the atmospheric parameters is as high as is their precision. We showed that while within one method the precision of the computations 
can be high, when results are compared to those obtained by means of other methods or from different data, various trends and offsets appear, proving that we are not yet able
to provide accurate values of $T_{\rm eff}$, $\log g$, and [Fe/H] for solar-type stars. 

KIC~5184732 is a good example of those difficulties. In Table~\ref{atmos}, we give the atmospheric parameters of that star derived independently from the 
spectra acquired with FRESCO, ESPaDOnS, and NARVAL. The atmospheric parameters computed with {\sf ARES+MOOG} from the ESPaDOnS and NARVAL data agree with each
other very well. A good consistency in the atmospheric parameters for all the spectra computed with ROTFIT is also found. However, the differences between those two sets of 
determinations amount to around 150~K in $T_{\rm eff}$, 0.12~dex in $\log g$, and 0.20~dex in [Fe/H]. For {\sf ROTFIT}, there are also less pronounced but still 
not negligible differences between $T_{\rm eff}$, $\log g$ and [Fe/H] derived from the observations acquired with FRESCO and those obtained with ESPaDOnS and NARVAL. 

The trends and discrepancies in the atmospheric parameters observed for stars hotter than 6,000~K represent yet another significant but not a new problem. The problem has 
been thoroughly discussed, but not solved, by \citet{torres2012}. Those authors compare atmospheric parameters obtained with {\sf SPC} and {\sf SME}, two codes in which 
the method of spectral synthesis is used, with the values of $T_{\rm eff}$, $\log g$, and [Fe/H] computed with {\sf MOOG}, which uses the curve-of-growth approach. The 
differences noticed by \citet{torres2012} are similar to those reported in the present paper. A similar trend can be noticed also in Fig.~3 $b$, in \citet{sousa2008}, 
where $T_{\rm eff}$ computed with {\sf ARES+MOOG} are compared with those obtained with {\sf SME}. The origin of those discrepancies is not clear but they seem to 
reflect systematic differences between atmospheric parameters obtained from the spectral synthesis and the analysis of equivalent widths. However, confirming that 
suspicion would require detailed examination of the input physics used in all the discussed methods which is beyond the scope of this paper.

The comparative analysis which we carried out showed that the currently available accuracy of atmospheric parameters of solar-type stars  
is $\pm 150$~K in $T_{\rm eff}$, $\pm 0.15$~dex in [Fe/H] and $\pm 0.3$~dex in $\log g$. This applies particularly to faint stars and those hotter than 
6,000~K. Since $\log g$ is the parameter most difficult to constrain in spectroscopic analysis, for stars showing solar-like pulsations and those with 
planetary transits, the seismic $\log g$ or the values of $\log g$ derived from the transit light curves may be used as an alternative value for asteroseismic modelling. 
Indeed, the $\log g$ values derived from transit light curves \citep{seager2003} are currently preferred in the investigation of the transiting planets 
\citep[c.f.][]{torres2012} whereas in asteroseismic modelling of stars showing solar-like oscillations, the seismic values of $\log g$ are preferred to the 
spectroscopic ones \citep[c.f.][]{morel2012}. Another potentially important application of the asteroseismically determined surface gravities
may be a validation test for the values of $\log g$ derived from the spectroscopic analyses \citep[see][]{creevey2013}.

\section{Summary}
\label{summary}
In this paper, we provided two sets of determinations of atmospheric parameters for 169 F, G, K, and M-type stars, dwarfs and giants, with $T_{\rm eff}$ 
ranging from 3,200 to 6,700~K. The first set was computed with ARES+MOOG, a method based on the analysis of the equivalent widths of spectral lines. The other was 
derived with ROTFIT, which makes use of the full spectrum by comparison to a grid of reference stars with well-known atmospheric parameters. The latter code was used
also to derive the projected rotational velocities of the stars. For 67 stars, the atmospheric parameters ($T_{\rm eff}$, $\log g$, and [Fe/H]) are provided for the 
first time. KIC~9693187 and KIC~11179629 are newly discovered double-lined spectroscopic binary systems.

The typical internal precision of the atmospheric parameters obtained with {\sf ARES+MOOG} is rather high: $\pm80$~K, $\pm 0.12$~dex, and $\pm 0.06$~dex in $T_{\rm eff}$, 
$\log g$, and [Fe/H], respectively. {\sf ROTFIT} displays a lower internal precision with typical errors of 110~K in $T_{\rm eff}$, and 0.21~dex in $\log g$ and [Fe/H],
however, the values of the atmospheric parameters are in good agreement with those derived by \citet{bruntt2012} and \citet{thygesen2012}. Therefore, we conclude that
both {\sf ROTFIT} and {\sf ARES+MOOG} produce determinations which can be safely used for asteroseismic modelling of stars or for studying stellar structure and evolution.

Having shown that for solar-type stars the present accuracy of the spectroscopic determinations of atmospheric parameters is $\pm150$~K in $T_{\rm eff}$, 
$\pm0.15$~dex in [Fe/H], and $\pm0.3$~dex in $\log g$, we emphasise the importance of collecting high-quality spectra with sufficiently large telescopes equipped 
with efficient  spectrographs. We stress also the need of examining reasons why for hot stars the spectral synthesis method and the curve-of-growth analysis yield 
atmospheric parameters which are systematically different.

\section*{Acknowledgements}
We thank Thierry Louge for help in retrieving the data from the CNRS/INSU CDAB/Bass2000 TBLegacy database operated by the University of Toulouse/OMP, Tarbes, France. 
We thank the Spanish Night time CAT for awarding the observing time to programs 61-Mercator3/11B, 119-NOT12/11A, and 61-NOT7/10A.
J.M.-\.Z., E.N., and D.D. acknowledge the Polish MNiSW grant N\,N203\,405139.
D.D. acknowledges the Polish National Science Centre grant No. 2011/01/N/ST9/00400.
A.F. acknowledges the Italian {\it Ministero dell'Istruzione, Universit\`a e  Ricerca} (MIUR).
S.G.S., A.M., and M.T. acknowledge the support of the European Research Council/European Community under the FP7 through Starting Grant agreement number 239953. 
S.G.S. and J.M.-\.Z. acknowledge the support from Funda\c{c}{\~a}o para a Ci{\^e}ncia e a Tecnologia (FCT) through the grant SFRH/BPD/47611/2008, the projects 
PTDC/CTE-AST/098528/2008, PTDC/CTE-AST/098754/2008, and the 'Coopera\c{c}{\~a}o Cientifica e Tecnologica FCT/Polonia 2011/2012 (Proc. 441.00 Polonia)', funded by FCT/MCTES, Portugal and POPH/FSE (EC). 
SB acknowledges funding from the European Research Council under the European Community's Seventh Framework Programme (FP7/2007--2013)/ERC grant agreement n$^\circ$227224 (PROSPERITY). 
KU acknowledges financial support by  the Spanish National Plan of R\&D for 2010, project AYA2010-17803. 
M.B. is F.R.S.-FNRS Postdoctoral Researcher, Belgium.

We thank the anonymous Referee for the comments which helped us improve this paper.

\begin{landscape}
\begin{table}
\begin{center}
\caption{The atmospheric parameters, the MK type, and the type of variability of our programme stars. In bold font, we indicate those stars for which the atmospheric parameters 
are derived for the first time. The symbols and acronyms used for different type of variability are explained in the footnote of this table.}
\label{atmos}
{\small
\begin{tabular}{rrrrrrrrrlrrrrrrrrll}
\hline\hline\noalign{\smallskip}
KIC\hspace{16pt} & $T_{\rm eff}$ & $\sigma$ & $\log g$ & $\sigma$ & [Fe/H] & $\sigma$ & $v\sin i$ & $\sigma$ & MK
                 & $T_{\rm eff}$ & $\sigma$ & $\log g$ & $\sigma$ & [Fe/H] & $\sigma$ & $\xi_{\rm t}$ & $\sigma$ & Instrument & var\\
         &\multicolumn{9}{c}{\hrulefill \,ROTFIT\,\hrulefill}
         &\multicolumn{8}{c}{\hrulefill \,ARES+MOOG\,\hrulefill} &\\
\hline
\hline
     1430163 & 6412 & 123 & 3.97 & 0.21 &$-$0.25 & 0.21 &  8.1 & 0.9 & F5IV    &  6833 &  87 & 4.70 & 0.11 &    0.02 & 0.06 &  2.12 &  0.10 &    NARVAL   &$\odot$ {\tiny (1)}       \\ 
     1435467 & 6169 & 130 & 3.95 & 0.21 &$-$0.04 & 0.22 &  9.0 & 1.0 & F8IV    &  6485 &  92 & 4.53 & 0.13 &    0.08 & 0.07 &  2.02 &  0.09 &    NARVAL   &$\odot$ {\tiny (1)}       \\ 
     2837475 & 6462 & 125 & 3.95 & 0.23 &$-$0.06 & 0.21 & 18.3 & 1.0 & F5IV-V  &  ---  & --- & ---  & ---  &    ---  & ---  &  ---  &  ---  &    ESPaDOnS &$\odot$ {\tiny (1)}       \\ 
\bf  3335176 & 3225 & 132 & 1.23 & 1.25 &$-$0.22 & 0.21 &  9.3 & 2.5 & M7II    &  ---  & --- & ---  & ---  &    ---  & ---  &  ---  &  ---  &\bf FIES     & PER {\tiny (2)}          \\ 
     3424541 & 6165 & 108 & 3.90 & 0.21 &   0.13 & 0.21 & 24.6 & 0.8 & G0IV    &  ---  & --- & ---  & ---  &    ---  & ---  &  ---  &  ---  &    NARVAL   &$\odot$ {\tiny (1)}       \\ 
     3427720 & 5949 &  98 & 4.26 & 0.21 &   0.00 & 0.21 &  2.0 & 0.7 & F9IV-V  &  6111 &  68 & 4.51 & 0.11 &    0.04 & 0.06 &  1.25 &  0.04 &    ESPaDOnS &$\odot$ {\tiny (1)}       \\ 
     3430868 & 4969 & 101 & 2.91 & 0.23 &$-$0.01 & 0.21 &  2.6 & 0.4 & G8III   &  5208 &  67 & 3.24 & 0.12 &    0.13 & 0.06 &  1.46 &  0.03 &    ESPaDOnS &                          \\ 
\bf  3443483 & 4856 &  93 & 3.05 & 0.21 &   0.04 & 0.21 & 11.1 & 0.2 & K1IV    &  5043 &  82 & 3.43 & 0.18 &    0.09 & 0.06 &  1.63 &  0.06 &\bf FIES     &$\odot$ {\tiny (1)}       \\ 
     3456181 & 6290 & 111 & 3.94 & 0.21 &$-$0.24 & 0.21 &  5.0 & 1.0 & F5IV-V  &  6584 &  91 & 4.43 & 0.11 & $-$0.02 & 0.07 &  2.01 &  0.11 &    NARVAL   &$\odot$ {\tiny (1)}       \\ 
     3632418 & 6148 & 111 & 3.94 & 0.21 &$-$0.19 & 0.21 &  6.3 & 0.5 & F6IV    &  6409 &  74 & 4.43 & 0.12 & $-$0.03 & 0.06 &  1.86 &  0.06 &    NARVAL   &$\odot\,\wp$ {\tiny (1,3)}\\ 
\bf  3643774 & 5928 &  96 & 4.26 & 0.22 &   0.17 & 0.21 &  1.4 & 1.4 & G2IV    &  6125 &  75 & 4.39 & 0.12 &    0.25 & 0.06 &  1.39 &  0.05 &\bf HERMES   &$\odot$ {\tiny (1)}       \\ 
\bf  3644223 & 4918 &  93 & 3.11 & 0.24 &$-$0.22 & 0.21 &  2.8 & 0.8 & G8III   &  ---  & --- & ---  & ---  &    ---  & ---  &  ---  &  ---  &\bf FRESCO   &$\odot$ {\tiny (4)}       \\ 
     3656476 & 5586 & 108 & 4.07 & 0.21 &   0.20 & 0.21 &  1.4 & 0.4 & G5IV    &  5719 &  64 & 4.26 & 0.11 &    0.28 & 0.05 &  1.11 &  0.03 &    ESPaDOnS &$\odot$ {\tiny (1)}       \\ 
     3733735 & 6548 & 156 & 3.99 & 0.22 &$-$0.12 & 0.21 & 13.0 & 1.4 & F5IV-V  &  ---  & --- & ---  & ---  &    ---  & ---  &  ---  &  ---  &    ESPaDOnS &$\odot$ {\tiny (1)}       \\ 
\bf  3747220 & 6668 & 147 & 4.18 & 0.21 &   0.00 & 0.21 & 50.8 &12.4 & F3V     &  ---  & --- & ---  & ---  &    ---  & ---  &  ---  &  ---  &\bf ESPaDOnS &                          \\ 
     4072740 & 4847 &  94 & 3.08 & 0.23 &   0.09 & 0.21 &  1.6 & 0.3 & K1IV    &  4960 &  77 & 3.49 & 0.13 &    0.19 & 0.06 &  1.13 &  0.06 &    NARVAL   &$\odot$ {\tiny (1)}       \\ 
\bf  4346201 & 6154 & 109 & 3.98 & 0.22 &$-$0.25 & 0.21 &  2.8 & 1.0 & F8V     &  6239 &  91 & 4.28 & 0.12 & $-$0.17 & 0.07 &  1.64 &  0.10 &\bf HERMES   &$\odot$ {\tiny (1)}       \\ 
     4586099 & 6304 & 109 & 3.92 & 0.21 &$-$0.20 & 0.21 &  2.3 & 0.7 & F5IV-V  &  6533 &  80 & 4.37 & 0.11 & $-$0.04 & 0.06 &  1.84 &  0.08 &    ESPaDOnS &$\odot$ {\tiny (1)}       \\ 
     4638884 & 6286 & 123 & 3.91 & 0.21 &$-$0.17 & 0.21 &  4.6 & 0.8 & F5IV-V  &  6684 &  98 & 4.58 & 0.17 & $-$0.05 & 0.08 &  3.39 &  0.28 &    NARVAL   &$\odot$ {\tiny (1)}       \\ 
\bf  4859338 & 6013 & 131 & 4.09 & 0.23 &   0.19 & 0.21 & 34.3 & 1.5 & G0IV    &  ---  & --- & ---  & ---  &  ---  & ---  &  ---  &  ---  &\bf HERMES   &$\odot$ {\tiny (13)}      \\ 
     4914923 & 5808 &  92 & 4.28 & 0.21 &   0.13 & 0.21 &  2.3 & 0.8 & G1.5V   &  5948 &  65 & 4.34 & 0.12 &    0.18 & 0.05 &  1.26 &  0.03 &    ESPaDOnS &$\odot$ {\tiny (1)}       \\ 
\bf  4931363 & 7045 & 128 & 4.07 & 0.22 &$-$0.05 & 0.21 & 65.9 & 8.0 & F0III   &  ---  & --- & ---  & ---  &  ---  & ---  &  ---  &  ---  &\bf ESPaDOnS &                          \\ 
\bf  4931390 & 6410 & 160 & 3.97 & 0.21 &$-$0.25 & 0.21 &  3.2 & 1.2 & F5IV-V  &  6862 &  80 & 4.55 & 0.11 & $-$0.02 & 0.06 &  1.93 &  0.09 &    ESPaDOnS &$\odot$ {\tiny (1)}       \\ 
     5021689 & 6141 & 107 & 3.94 & 0.21 &$-$0.16 & 0.22 &  7.0 & 0.6 & F8IV    &  6378 &  80 & 4.55 & 0.13 & $-$0.02 & 0.06 &  1.90 &  0.08 &    ESPaDOnS &$\odot$ {\tiny (1)}       \\ 
\bf  5024851 & 4046 &  92 & 1.77 & 0.21 &$-$0.18 & 0.21 &  1.9 & 0.7 & K4III   &  ---  & --- & ---  & ---  &  ---  & ---  &  ---  &  ---  &\bf ESPaDOnS &$\odot$ {\tiny (5)}       \\ 
\bf  5080290 & 5157 & 169 & 3.60 & 0.35 &$-$0.06 & 0.22 &  4.6 & 0.9 & K0III-IV&  5072 &  77 & 3.31 & 0.16 & $-$0.10 & 0.07 &  0.69 &  0.07 &\bf HERMES   &$\delta$ Sct {\tiny (6)}  \\ 
     ...~~~~~& 5261 & 182 & 4.21 & 0.26 &   0.01 & 0.23 &  6.1 & 0.5 & K0III-IV&  5064 &  78 & 3.28 & 0.13 & $-$0.14 & 0.06 &  0.79 &  0.06 &\bf ESPaDOnS &                          \\ 
\bf  5112786 & 4207 &  92 & 1.99 & 0.21 &$-$0.17 & 0.21 &  2.5 & 0.9 & K3III   &  4477 & 114 & 2.21 & 0.22 & $-$0.13 & 0.07 &  1.83 &  0.08 &\bf ESPaDOnS &                          \\ 
     5184732 & 5669 &  97 & 4.07 & 0.21 &   0.24 & 0.21 &  2.8 & 0.3 & G4V     &  ---  & --- & ---  & ---  &  ---  & ---  &  ---  &  ---  &\bf FRESCO   &$\odot$ {\tiny (1)}       \\ 
     ...~~~~~& 5723 & 103 & 4.18 & 0.23 &   0.21 & 0.21 &  2.2 & 0.6 & G4V     &  5894 &  68 & 4.31 & 0.12 &    0.43 & 0.06 &  1.18 &  0.03 &    ESPaDOnS$^*$&                       \\ 
     ...~~~~~& 5740 & 115 & 4.22 & 0.22 &   0.18 & 0.21 &  2.4 & 0.5 & G1V     &  5877 &  68 & 4.34 & 0.11 &    0.40 & 0.06 &  1.14 &  0.03 &    NARVAL$^*$&                         \\ 
\bf  5199859 & 3722 & 133 & 1.63 & 0.35 &$-$0.07 & 0.21 & 10.8 & 1.3 & M0III   &  ---  & --- & ---  & ---  &  ---  & ---  &  ---  &  ---  &\bf FIES     &                          \\ 
     5371516 & 6138 &  90 & 3.98 & 0.22 &   0.10 & 0.21 &  9.7 & 1.2 & F8IV    &  6526 & 107 & 4.49 & 0.15 &    0.11 & 0.08 &  2.35 &  0.14 &    ESPaDOnS &$\odot$ {\tiny (1)}       \\ 
     5450445 & 6099 &  99 & 4.13 & 0.21 &   0.05 & 0.21 &  5.4 & 0.6 & F8V     &  6396 &  75 & 4.49 & 0.11 &    0.23 & 0.06 &  1.75 &  0.06 &    NARVAL   &$\odot$ {\tiny (1)}       \\ 
     5512589 & 5764 &  95 & 4.22 & 0.21 &   0.06 & 0.21 &  1.6 & 0.4 & G3V     &  5812 &  66 & 4.05 & 0.11 &    0.12 & 0.06 &  1.20 &  0.03 &    NARVAL   &$\odot$ {\tiny (1)}       \\ 
\bf  5557932 & 5936 & 100 & 4.37 & 0.21 &   0.00 & 0.21 & 13.7 & 0.3 & G1.5V   &  ---  & --- & ---  & ---  &  ---  & ---  &  ---  &  ---  &\bf ESPaDOnS &RS\,CVn {\tiny (7)}       \\ 
     5596656 & 5375 & 112 & 3.99 & 0.24 &$-$0.18 & 0.21 &  3.8 & 0.4 & G5IV    &  5188 &  69 & 3.75 & 0.13 & $-$0.44 & 0.06 &  1.05 &  0.05 &    ESPaDOnS &$\odot$ {\tiny (1)}       \\ 
\bf  5620305 & 5190 & 148 & 3.49 & 0.32 &$-$0.01 & 0.21 &  4.2 & 1.1 & K0III-IV&  5040 &  70 & 2.95 & 0.12 & $-$0.01 & 0.06 &  0.51 &  0.05 &\bf HERMES   &                          \\ 
     5701829 & 4927 & 104 & 3.19 & 0.22 &$-$0.24 & 0.21 &  2.3 & 0.5 & K0IV    &  ---  & --- & ---  & ---  &  ---  & ---  &  ---  &  ---  &\bf FRESCO   &$\odot$ {\tiny (1)}       \\ 
     ...~~~~~& 4914 &  92 & 3.18 & 0.22 &$-$0.13 & 0.21 &  2.4 & 0.7 & K0IV    &  4962 &  69 & 3.39 & 0.13 & $-$0.17 & 0.06 &  1.13 &  0.04 &    ESPaDOnS &                          \\ 
\bf  5737655 & 5163 & 101 & 2.88 & 0.25 &$-$0.44 & 0.21 &  3.8 & 0.6 & G4III-IV&  5121 &  63 & 2.83 & 0.10 & $-$0.56 & 0.05 &  1.68 &  0.02 &\bf ESPaDOnS &$\odot$ {\tiny (1)}       \\ 
     5773345 & 6007 & 112 & 4.17 & 0.21 &   0.13 & 0.21 &  3.4 & 1.1 & G0.5IV  &  6399 &  71 & 4.36 & 0.11 &    0.30 & 0.06 &  1.92 &  0.05 &    ESPaDOnS &$\odot$ {\tiny (1)}       \\ 
\hline                                     
\end{tabular}                              
}                                          
\end{center}                               
\end{table}
\end{landscape}

\setcounter{table}{2}

\begin{landscape}
\begin{table}
\begin{center}
\caption{continuation.}
{\small
\begin{tabular}{rrrrrrrrrlrrrrrrrrll}
\hline\hline\noalign{\smallskip}
KIC\hspace{16pt}     & $T_{\rm eff}$ & $\sigma$ & $\log g$ & $\sigma$ & [Fe/H] & $\sigma$ & $v\sin i$ & $\sigma$ & MK
                     & $T_{\rm eff}$ & $\sigma$ & $\log g$ & $\sigma$ & [Fe/H] & $\sigma$ & $\xi_{\rm t}$ & $\sigma$ & Instrument & var\\
		 &\multicolumn{9}{c}{\hrulefill \,ROTFIT\,\hrulefill}
		 &\multicolumn{8}{c}{\hrulefill \,ARES+MOOG\,\hrulefill} &\\
\hline
\hline
     5774694 & 5804 &  91 & 4.34 & 0.21 &   0.08 & 0.21 &  3.6 & 0.5 & G2V     &  5923 &  65 & 4.56 & 0.10 &    0.10 & 0.05 &  1.17 &  0.03 &    ESPaDOnS$^*$&$\odot$ {\tiny (1)}\\ 
     ...~~~~~& 5801 &  98 & 4.34 & 0.21 &   0.06 & 0.21 &  3.6 & 0.6 & G3V     &  5950 &  64 & 4.58 & 0.10 &    0.09 & 0.05 &  1.19 &  0.03 &    NARVAL$^*$&                     \\ 
\bf  5952403 & 5058 & 106 & 2.99 & 0.26 &   0.01 & 0.21 & 13.6 & 0.1 & G8III   &  ---  & --- & ---  & ---  &    ---  & ---  &  ---  &  ---  &\bf FIES                            \\ 
     5955122 & 5952 & 100 & 4.13 & 0.21 &$-$0.05 & 0.22 &  4.5 & 0.6 & F9IV-V  &  6092 &  69 & 4.26 & 0.12 & $-$0.06 & 0.06 &  1.66 &  0.05 &    ESPaDOnS &$\odot$ {\tiny (1)}   \\ 
     6116048 & 5991 & 124 & 4.09 & 0.22 &$-$0.24 & 0.23 &  2.9 & 0.6 & F9IV-V  &  6152 &  66 & 4.53 & 0.10 & $-$0.14 & 0.05 &  1.36 &  0.04 &    ESPaDOnS &$\odot$ {\tiny (1)}   \\ 
     6225718 & 6138 & 106 & 3.96 & 0.21 &$-$0.23 & 0.22 &  2.4 & 0.5 & F8V     &  6366 &  70 & 4.61 & 0.11 & $-$0.07 & 0.06 &  1.50 &  0.05 &    NARVAL   &$\odot$ {\tiny (1)}   \\ 
\bf  6285677 & 5849 &  97 & 4.32 & 0.22 &   0.06 & 0.22 &  7.6 & 1.0 & G2V     &  6205 &  73 & 4.48 & 0.11 &    0.23 & 0.06 &  1.48 &  0.05 &\bf HERMES                          \\ 
     ...~~~~~& 5907 &  94 & 4.18 & 0.21 &   0.02 & 0.21 &  7.8 & 0.9 & G0.5IV  &  ---  & --- & ---  & ---  &    ---  & ---  &  ---  &  ---  &\bf FRESCO                          \\ 
\bf  6370489 & 6241 & 116 & 3.98 & 0.21 &$-$0.35 & 0.21 &  4.4 & 0.8 & F8V     &  ---  & --- & ---  & ---  &    ---  & ---  &  ---  &  ---  &\bf FIES     &$\odot$ {\tiny (1)}   \\ 
     6442183 & 5736 &  96 & 4.26 & 0.21 &$-$0.07 & 0.21 &  1.7 & 0.5 & G1V     &  5738 &  62 & 4.14 & 0.10 & $-$0.12 & 0.05 &  1.15 &  0.02 &    NARVAL   &$\odot$ {\tiny (1)}   \\ 
     6508366 & 6332 & 117 & 3.91 & 0.21 &$-$0.07 & 0.21 & 18.0 & 1.0 & F6IV    &  ---  & --- & ---  & ---  &    ---  & ---  &  ---  &  ---  &    ESPaDOnS &$\odot$ {\tiny (1)}   \\ 
\bf  6590668 & 4463 &  93 & 2.02 & 0.22 &$-$0.22 & 0.21 &  4.0 & 1.1 & K1III   &  ---  & --- & ---  & ---  &    ---  & ---  &  ---  &  ---  &\bf FRESCO                          \\ 
     6603624 & 5471 & 128 & 4.02 & 0.23 &   0.17 & 0.21 &  1.4 & 0.7 & G8IV-V  &  5718 &  78 & 4.44 & 0.13 &    0.28 & 0.06 &  1.16 &  0.06 &    ESPaDOnS &$\odot$ {\tiny (1)}   \\ 
     6679371 & 6344 & 131 & 3.92 & 0.21 &$-$0.10 & 0.21 & 11.0 & 1.0 & F5IV-V  &  ---  & --- & ---  & ---  &    ---  & ---  &  ---  &  ---  &    NARVAL   &$\odot$ {\tiny (1)}   \\ 
\bf  6766118 & 4892 &  93 & 2.73 & 0.21 &   0.05 & 0.21 &  2.7 & 0.6 & K0III   &  ---  & --- & ---  & ---  &    ---  & ---  &  ---  &  ---  &\bf FRESCO                          \\ 
     6933899 & 5837 &  97 & 4.21 & 0.22 &   0.04 & 0.21 &  2.0 & 0.6 & G0.5IV  &  5921 &  65 & 4.12 & 0.11 &    0.04 & 0.06 &  1.29 &  0.03 &    NARVAL   &$\odot$ {\tiny (1)}   \\ 
     7103006 & 6180 & 120 & 3.92 & 0.21 &$-$0.07 & 0.22 &  8.9 & 0.6 & F8IV    &  6685 &  86 & 4.50 & 0.11 &    0.19 & 0.06 &  1.98 &  0.08 &    NARVAL   &$\odot$ {\tiny (1)}   \\ 
     7206837 & 6142 & 112 & 4.05 & 0.21 &   0.05 & 0.21 &  6.7 & 0.5 & F8IV    &  6573 &  80 & 4.61 & 0.11 &    0.22 & 0.06 &  1.93 &  0.06 &    NARVAL   &$\odot$ {\tiny (1)}   \\ 
     7282890 & 6207 &  97 & 3.89 & 0.22 &   0.02 & 0.21 & 21.0 & 1.0 & F6IV    &  ---  & --- & ---  & ---  &    ---  & ---  &  ---  &  ---  &    ESPaDOnS &$\odot$ {\tiny (1)}   \\ 
     7510397 & 6120 &  97 & 3.94 & 0.21 &$-$0.26 & 0.22 &  2.2 & 0.8 & F6IV    &  6362 &  80 & 4.54 & 0.12 & $-$0.08 & 0.06 &  1.66 &  0.07 &    ESPaDOnS &$\odot$ {\tiny (1)}   \\ 
     7529180 & 6470 & 128 & 4.03 & 0.21 &$-$0.06 & 0.21 & 27.0 & 1.7 & F5IV-V  &  ---  & --- & ---  & ---  &    ---  & ---  &  ---  &  ---  &    NARVAL   &$\odot$ {\tiny (1)}   \\ 
     7662428 & 6143 &  97 & 4.03 & 0.21 &   0.10 & 0.21 &  9.3 & 0.8 & F8V     &  6504 & 141 & 4.93 & 0.19 & $-$0.09 & 0.10 &  1.58 &  0.22 &    ESPaDOnS &$\odot$ {\tiny (13)}  \\ 
     7668623 & 6159 & 105 & 3.94 & 0.21 &$-$0.10 & 0.23 &  7.6 & 0.7 & F8IV    &  6580 & 112 & 4.56 & 0.15 &    0.03 & 0.08 &  2.54 &  0.21 &    ESPaDOnS &$\odot$ {\tiny (1)}   \\ 
     7680114 & 5799 &  91 & 4.25 & 0.21 &   0.08 & 0.21 &  1.4 & 0.8 & G0V     &  5955 &  68 & 4.41 & 0.11 &    0.12 & 0.06 &  1.30 &  0.04 &    NARVAL   &$\odot$ {\tiny (1)}   \\ 
\bf  7730305 & 6060 & 104 & 4.25 & 0.22 &   0.09 & 0.21 & 12.6 & 1.1 & F8V     &  6304 &  81 & 4.67 & 0.11 &    0.17 & 0.06 &  1.69 &  0.07 &\bf HERMES                          \\ 
     ...~~~~~& 6030 & 104 & 4.17 & 0.21 &   0.01 & 0.21 & 15.0 & 0.8 & F8V     &  ---  & --- & ---  & ---  &    ---  & ---  &  ---  &  ---  &\bf FRESCO                          \\ 
     7747078 & 5994 & 113 & 4.04 & 0.23 &$-$0.19 & 0.23 &  3.8 & 0.8 & F9IV-V  &  6114 &  78 & 4.37 & 0.12 & $-$0.11 & 0.06 &  1.65 &  0.07 &    ESPaDOnS &$\odot$ {\tiny (1)}   \\ 
     7799349 & 4954 &  92 & 3.33 & 0.22 &   0.14 & 0.21 &  1.1 & 0.4 & K1IV    &  5175 &  84 & 3.81 & 0.15 &    0.24 & 0.07 &  1.31 &  0.07 &    NARVAL   &$\odot$ {\tiny (1)}   \\ 
\bf  7799575 & 3941 &  92 & 1.69 & 0.21 &$-$0.17 & 0.21 &  2.2 & 0.7 & K5III   &  ---  & --- & ---  & ---  &    ---  & ---  &  ---  &  ---  &\bf ESPaDOnS &Mira {\tiny (8)}      \\ 
     7800289 & 6398 & 133 & 3.96 & 0.21 &$-$0.17 & 0.21 & 18.6 & 1.1 & F5IV    &  ---  & --- & ---  & ---  &    ---  & ---  &  ---  &  ---  &    NARVAL   &$\odot$ {\tiny (9)}   \\ 
     7871531 & 5498 & 117 & 4.31 & 0.21 &$-$0.12 & 0.21 &  2.2 & 0.8 & G5V     &  5461 &  67 & 4.40 & 0.12 & $-$0.26 & 0.06 &  0.87 &  0.05 &    ESPaDOnS &$\odot$ {\tiny (1)}   \\ 
     7940546 & 6243 & 101 & 3.92 & 0.21 &$-$0.25 & 0.21 &  6.6 & 0.8 & F6IV    &  6427 &  82 & 4.52 & 0.12 & $-$0.11 & 0.06 &  2.09 &  0.09 &    ESPaDOnS$^*$&$\odot$ {\tiny (1)}\\ 
     ...~~~~~& 6226 & 119 & 3.94 & 0.21 &$-$0.24 & 0.21 &  7.0 & 0.7 & F6IV    &  6472 &  84 & 4.59 & 0.12 & $-$0.11 & 0.06 &  2.32 &  0.12 &    NARVAL$^*$&                     \\ 
     7970740 & 5354 & 111 & 4.36 & 0.21 &$-$0.31 & 0.21 &  2.4 & 0.5 & G9V     &  5287 &  68 & 4.49 & 0.11 & $-$0.52 & 0.05 &  0.59 &  0.08 &    ESPaDOnS &$\odot$ {\tiny (1)}   \\ 
     7976303 & 6119 & 106 & 3.97 & 0.21 &$-$0.38 & 0.21 &  3.1 & 0.5 & F8V     &  6203 &  76 & 4.15 & 0.11 & $-$0.41 & 0.06 &  1.62 &  0.07 &    ESPaDOnS &$\odot$ {\tiny (1)}   \\ 
     7985370 & 5836 & 103 & 4.39 & 0.21 &   0.02 & 0.21 & 16.4 & 0.3 & G1.5V   &  ---  & --- & ---  & ---  &    ---  & ---  &  ---  &  ---  &\bf HERMES   &PER {\tiny (2)}       \\ 
     ...~~~~~& 5849 &  90 & 4.28 & 0.21 &$-$0.10 & 0.21 & 17.3 & 0.4 & G1.5V   &  ---  & --- & ---  & ---  &    ---  & ---  &  ---  &  ---  &\bf FRESCO   &rot/act {\tiny (6,11)}   \\ 
     8006161 & 5258 &  97 & 4.13 & 0.25 &   0.23 & 0.21 &  2.0 & 0.5 & G8V     &  5431 &  82 & 4.45 & 0.13 &    0.30 & 0.06 &  0.95 &  0.10 &    ESPaDOnS$^*$&$\odot$ {\tiny (1)}\\ 
     ...~~~~~& 5211 & 101 & 4.05 & 0.24 &   0.20 & 0.21 &  1.9 & 0.3 & G8V     &  5468 &  77 & 4.41 & 0.13 &    0.29 & 0.06 &  1.07 &  0.07 &    NARVAL$^*$&                     \\ 
     8026226 & 6276 &  94 & 3.90 & 0.21 &$-$0.20 & 0.21 &  7.4 & 0.5 & F5IV-V  &  6469 &  78 & 4.32 & 0.13 & $-$0.13 & 0.06 &  2.72 &  0.18 &    ESPaDOnS &$\odot$ {\tiny (9)}   \\ 
     8179536 & 6160 & 112 & 3.98 & 0.21 &$-$0.16 & 0.21 &  8.1 & 0.6 & F6IV    &  6536 &  74 & 4.64 & 0.11 &    0.13 & 0.06 &  1.61 &  0.05 &    NARVAL   &$\odot$ {\tiny (1)}   \\ 
\hline                                     
\end{tabular}                              
}                                          
\end{center}                               
\end{table}
\end{landscape}

\setcounter{table}{2}

\begin{landscape}
\begin{table}
\begin{center}
\caption{continuation.}
{\small
\begin{tabular}{rrrrrrrrrlrrrrrrrrll}
\hline\hline\noalign{\smallskip}
KIC\hspace{16pt}     & $T_{\rm eff}$ & $\sigma$ & $\log g$ & $\sigma$ & [Fe/H] & $\sigma$ & $v\sin i$ & $\sigma$ & MK
                     & $T_{\rm eff}$ & $\sigma$ & $\log g$ & $\sigma$ & [Fe/H] & $\sigma$ & $\xi_{\rm t}$ & $\sigma$ & Instrument & var\\
         &\multicolumn{9}{c}{\hrulefill \,ROTFIT\,\hrulefill}
         &\multicolumn{8}{c}{\hrulefill \,ARES+MOOG\,\hrulefill} &\\
\hline
\hline
     8211551 & 4812 &  93 & 2.83 & 0.23 &$-$0.12 & 0.21 &  1.9 & 0.5 & G9III   &  4882 &  68 & 2.76 & 0.12 & $-$0.15 & 0.06 &  1.54 &  0.03 &    ESPaDOnS$^*$                          \\ 
     ...~~~~~& 4820 &  93 & 2.83 & 0.22 &$-$0.10 & 0.21 &  2.0 & 0.3 & G9III   &  4887 &  70 & 2.69 & 0.13 & $-$0.17 & 0.06 &  1.56 &  0.03 &    NARVAL$^*$                            \\ 
     8228742 & 6061 & 108 & 4.02 & 0.22 &$-$0.12 & 0.21 &  3.3 & 1.1 & F9IV-V  &  6295 &  76 & 4.42 & 0.11 &    0.00 & 0.06 &  1.71 &  0.06 &    ESPaDOnS &$\odot$ {\tiny (1)}         \\ 
\bf  8343931 & 6506 & 125 & 4.09 & 0.22 &$-$0.03 & 0.21 & 43.2 & 4.0 & F5IV-V  &  ---  & --- & ---  & ---  &    ---  & ---  &  ---  &  ---  &\bf ESPaDOnS                              \\ 
\bf  8346342 & 6141 & 119 & 3.93 & 0.21 &$-$0.05 & 0.22 &  6.9 & 0.8 & F8IV    &  6573 & 139 & 4.59 & 0.12 &    0.21 & 0.10 &  1.87 &  0.15 &\bf ESPaDOnS                              \\ 
\bf  8352528 & 3972 &  89 & 1.69 & 0.21 &$-$0.18 & 0.21 &  2.2 & 0.9 & K5III   &  ---  & --- & ---  & ---  &    ---  & ---  &  ---  &  ---  &\bf ESPaDOnS                              \\ 
     8360349 & 6176 &  91 & 3.92 & 0.21 &   0.07 & 0.21 & 10.6 & 0.7 & F8IV    &  6762 & 156 & 4.92 & 0.15 &    0.07 & 0.10 &  3.45 &  0.37 &    ESPaDOnS &$\odot$ {\tiny (1)}         \\ 
     8367710 & 6227 & 116 & 3.92 & 0.21 &   0.02 & 0.21 & 15.0 & 1.1 & F6IV    &  ---  & --- & ---  & ---  &    ---  & ---  &  ---  &  ---  &    ESPaDOnS &$\odot$ {\tiny (1)}         \\ 
     8379927 & 5998 & 108 & 4.25 & 0.21 &$-$0.03 & 0.22 &  8.8 & 0.8 & F9IV-V  &  6225 &  95 & 4.76 & 0.13 & $-$0.23 & 0.07 &  2.01 &  0.13 &    ESPaDOnS$^*$ &$\odot$ {\tiny (1)}     \\ 
     ...~~~~~& 6000 & 112 & 4.12 & 0.22 &$-$0.05 & 0.23 & 13.0 & 2.0 & F9IV-V  &  6202 &  73 & 4.47 & 0.12 & $-$0.20 & 0.06 &  0.95 &  0.05 &    NARVAL$^*$                            \\ 
     8394589 & 6111 & 116 & 3.98 & 0.21 &$-$0.37 & 0.21 &  4.4 & 0.8 & F8V     &  6231 &  75 & 4.54 & 0.11 & $-$0.24 & 0.06 &  1.36 &  0.07 &    NARVAL   &$\odot$ {\tiny (1)}         \\ 
     8429280 & 5029 & 103 & 4.35 & 0.21 &$-$0.04 & 0.21 & 34.8 & 0.6 & K2V     &  ---  & --- & ---  & ---  &    ---  & ---  &  ---  &  ---  &\bf FRESCO   &rot/act {\tiny (12)}        \\ 
     ...~~~~~& 5108 & 114 & 4.56 & 0.23 &   0.06 & 0.21 & 33.2 & 1.0 & K1V     &  ---  & --- & ---  & ---  &    ---  & ---  &  ---  &  ---  &\bf HERMES                                \\ 
     8491147 & 5007 &  95 & 2.92 & 0.24 &$-$0.24 & 0.21 &  2.5 & 0.6 & G8III   &  5065 &  65 & 2.75 & 0.12 & $-$0.31 & 0.06 &  1.57 &  0.02 &    ESPaDOnS &$\odot$ {\tiny (4)}         \\ 
     8524425 & 5671 & 105 & 4.17 & 0.22 &   0.12 & 0.21 &  1.1 & 0.5 & G2.5V   &  5664 &  65 & 4.09 & 0.11 &    0.13 & 0.05 &  1.16 &  0.03 &    NARVAL   &$\odot$ {\tiny (1)}         \\ 
     8542853 & 5594 &  99 & 4.34 & 0.21 &$-$0.09 & 0.21 &  2.1 & 0.6 & G6V     &  5580 &  68 & 4.54 & 0.12 & $-$0.20 & 0.06 &  0.85 &  0.06 &    ESPaDOnS                              \\ 
     8547390 & 4732 &  90 & 2.80 & 0.21 &$-$0.01 & 0.21 &  3.0 & 0.3 & K0III   &  4870 &  74 & 2.86 & 0.15 &    0.12 & 0.06 &  1.60 &  0.04 &    ESPaDOnS &$\odot$ {\tiny (1)}         \\ 
     8561221 & 5290 & 115 & 3.76 & 0.23 &$-$0.04 & 0.21 &  1.9 & 0.6 & G9.5IV  &  5352 &  68 & 3.80 & 0.11 & $-$0.04 & 0.06 &  1.14 &  0.04 &    NARVAL   &$\odot$ {\tiny (1)}         \\ 
     8579578 & 6297 & 144 & 3.91 & 0.21 &$-$0.06 & 0.21 & 19.3 & 1.0 & F6IV    &  ---  & --- & ---  & ---  &    ---  & ---  &  ---  &  ---  &    NARVAL   &$\odot$ {\tiny (1)}         \\ 
\bf  8677933 & 5946 & 161 & 3.92 & 0.29 &   0.15 & 0.22 & 49.6 & 0.7 & G0IV    &  ---  & --- & ---  & ---  &    ---  & ---  &  ---  &  ---  &\bf ESPaDOnS &$\odot$ {\tiny (1)}         \\ 
     8694723 & 6258 & 117 & 3.97 & 0.21 &$-$0.42 & 0.21 &  4.6 & 1.0 & G0IV    &  6445 &  80 & 4.55 & 0.11 & $-$0.39 & 0.06 &  1.91 &  0.11 &    NARVAL   &$\odot$ {\tiny (1)}         \\ 
     ...~~~~~& 6287 & 116 & 4.00 & 0.21 &$-$0.38 & 0.22 &  3.8 & 0.7 & G0IV    &  6489 &  85 & 4.50 & 0.13 & $-$0.35 & 0.06 &  1.98 &  0.13 &\bf FIES                                  \\ 
     8702606 & 5621 & 106 & 4.08 & 0.21 &   0.00 & 0.21 &  0.7 & 0.7 & G5IV-V  &  5578 &  62 & 3.89 & 0.10 & $-$0.06 & 0.05 &  1.16 &  0.02 &    ESPaDOnS &$\odot$ {\tiny (1)}         \\ 
     8738809 & 6039 & 104 & 4.19 & 0.21 &   0.07 & 0.21 &  2.2 & 0.9 & G0.5IV  &  6207 &  68 & 4.17 & 0.11 &    0.12 & 0.06 &  1.65 &  0.03 &    NARVAL   &$\odot$ {\tiny (1)}         \\ 
     8751420 & 5281 & 115 & 3.86 & 0.24 &$-$0.11 & 0.21 &  1.1 & 0.5 & G8IV    &  5330 &  62 & 3.84 & 0.10 & $-$0.14 & 0.05 &  1.07 &  0.02 &    NARVAL   &$\odot$ {\tiny (1)}         \\ 
     8760414 & 5850 & 166 & 3.94 & 0.26 &$-$0.90 & 0.29 &  3.4 & 2.3 & G0IV    &  5924 &  77 & 4.53 & 0.11 & $-$1.00 & 0.06 &  1.38 &  0.11 &    NARVAL   &$\odot$ {\tiny (1)}         \\ 
\bf  8816903 & 7063 & 142 & 4.12 & 0.21 &$-$0.05 & 0.21 & 57.6 & 5.0 & F0V     &  ---  & --- & ---  & ---  &    ---  & ---  &  ---  &  ---  &\bf ESPaDOnS                              \\ 
\bf  8831759 & 3877 & 107 & 1.66 & 0.24 &$-$0.11 & 0.21 &  2.4 & 0.7 & M1III   &  4920 & 209 & 3.94 & 0.34 & $-$0.14 & 0.10 &  3.65 &  0.58 &\bf ESPaDOnS                              \\ 
\bf  8866102 & 6195 & 134 & 3.95 & 0.21 &$-$0.16 & 0.21 & 11.0 & 0.8 & F6IV    &  ---  & --- & ---  & ---  &    ---  & ---  &  ---  &  ---  &\bf ESPaDOnS &$\wp?$ {\tiny (10)}         \\ 
     8938364 & 5702 & 101 & 4.25 & 0.21 &$-$0.16 & 0.22 &  2.0 & 0.9 & G3V     &  5808 &  71 & 4.31 & 0.12 & $-$0.10 & 0.06 &  1.10 &  0.05 &    NARVAL   &$\odot$ {\tiny (1)}         \\ 
     9025370 &  --- & --- &  --- &  --- &    --- &  --- &  --- & --- & ---     &  ---  & --- & ---  & ---  &    ---  & ---  &  ---  &  ---  &    ESPaDOnS &$\odot$\,SB2 {\tiny (1,14)}   \\ 
     9098294 & 5766 &  96 & 4.27 & 0.21 &$-$0.22 & 0.22 &  2.6 & 0.6 & G3V     &  5959 &  80 & 4.56 & 0.12 & $-$0.04 & 0.06 &  1.13 &  0.07 &    NARVAL   &$\odot$      {\tiny (1)}    \\ 
\bf  9116461 & 6358 & 108 & 3.95 & 0.21 &$-$0.14 & 0.21 & 14.1 & 0.6 & F5IV-V  &  ---  & --- & ---  & ---  &    ---  & ---  &  ---  &  ---  &\bf ESPaDOnS &$\odot$ {\tiny (1)}         \\ 
     9139151 & 6004 &  94 & 4.26 & 0.21 &   0.07 & 0.21 &  3.2 & 0.5 & G0.5IV  &  6213 &  67 & 4.64 & 0.11 &    0.17 & 0.06 &  1.24 &  0.04 &    ESPaDOnS &$\odot$ {\tiny (1)}         \\ 
     9139163 & 6175 & 123 & 3.99 & 0.22 &   0.00 & 0.21 &  2.0 & 1.0 & F8IV    &  6577 &  69 & 4.44 & 0.10 &    0.21 & 0.06 &  1.68 &  0.04 &    ESPaDOnS$^*$ &$\odot$ {\tiny (1)}     \\ 
     ...~~~~~& 6151 & 128 & 3.98 & 0.21 &$-$0.05 & 0.22 &  1.9 & 0.8 & F8IV    &  6584 &  67 & 4.47 & 0.11 &    0.19 & 0.05 &  1.70 &  0.03 &    NARVAL$^*$                            \\ 
     9206432 & 6204 & 142 & 3.95 & 0.21 &$-$0.02 & 0.22 &  1.7 & 1.2 & F8IV    &  6772 &  73 & 4.61 & 0.11 &    0.28 & 0.06 &  1.92 &  0.05 &    ESPaDOnS &$\odot$ {\tiny (1)}         \\ 
     9226926 & 6580 & 142 & 4.12 & 0.21 &$-$0.15 & 0.22 & 30.8 & 3.0 & F5V     &  ---  & --- & ---  & ---  &    ---  & ---  &  ---  &  ---  &    NARVAL   &$\odot$ {\tiny (1)}         \\ 
\bf  9289275 & 5931 & 103 & 4.25 & 0.22 &   0.07 & 0.22 &  2.7 & 1.5 & G0.5IV  &  6208 &  77 & 4.40 & 0.12 &    0.20 & 0.06 &  1.51 &  0.06 &\bf HERMES   &$\odot$ {\tiny (1)}         \\ 
\bf  9414417 & 6242 & 104 & 3.92 & 0.21 &$-$0.19 & 0.21 &  6.0 & 1.1 & F6IV    &  6496 & 124 & 4.66 & 0.13 & $-$0.07 & 0.09 &  2.55 &  0.26 &\bf HERMES   &$\odot\,\wp?$ {\tiny (1,10)}\\ 
\bf  9512063 & 5882 & 112 & 4.14 & 0.22 &$-$0.19 & 0.24 &  2.5 & 1.3 & F9IV-V  &  5842 &  72 & 3.87 & 0.11 & $-$0.15 & 0.06 &  1.12 &  0.04 &\bf HERMES   &$\odot$ {\tiny (1)}         \\ 
\bf  9514879 & 5971 &  92 & 4.31 & 0.21 &   0.02 & 0.21 & 10.1 & 0.3 & G1.5V   &  6190 &  79 & 4.70 & 0.12 &    0.12 & 0.06 &  1.60 &  0.07 &\bf FIES                                  \\ 
     9532030 & 4472 &  92 & 2.35 & 0.22 &$-$0.11 & 0.21 &  3.6 & 0.5 & G9III   &  4596 &  85 & 2.53 & 0.17 & $-$0.06 & 0.06 &  1.74 &  0.06 &    ESPaDOnS                              \\ 
\hline                                                                                   
\end{tabular}                                                                            
}
\end{center}
\end{table}
\end{landscape}

\setcounter{table}{2}

\begin{landscape}
\begin{table}
\begin{center}
\caption{continuation.}
{\small
\begin{tabular}{rrrrrrrrrlrrrrrrrrll}
\hline\hline\noalign{\smallskip}
KIC\hspace{16pt}     & $T_{\rm eff}$ & $\sigma$ & $\log g$ & $\sigma$ & [Fe/H] & $\sigma$ & $v\sin i$ & $\sigma$ & MK
                     & $T_{\rm eff}$ & $\sigma$ & $\log g$ & $\sigma$ & [Fe/H] & $\sigma$ & $\xi_{\rm t}$ & $\sigma$ & Instrument & var\\
         &\multicolumn{9}{c}{\hrulefill \,ROTFIT\,\hrulefill}
         &\multicolumn{8}{c}{\hrulefill \,ARES+MOOG\,\hrulefill} &\\
\hline
\hline
\bf  9534041 & 5061 &  96 & 3.10 & 0.23 &   0.02 & 0.21 &  3.2 & 0.6 & G8III   &  5278 &  72 & 3.28 & 0.12 & $-$0.01 & 0.06 &  1.49 &  0.04 &\bf ESPaDOnS &$\odot$ {\tiny (5)}         \\ 
\bf  9605196 & 4455 &  90 & 1.91 & 0.23 &$-$0.20 & 0.21 &  3.5 & 0.8 & K1III   &  ---  & --- & ---  & ---  &    ---  & ---  &  ---  &  ---  &\bf FRESCO                                \\ 
\bf  9655101 & 5039 & 129 & 3.02 & 0.26 &   0.00 & 0.21 &  3.5 & 0.7 & G8III   &  5227 &  73 & 3.31 & 0.13 & $-$0.02 & 0.06 &  1.53 &  0.04 &\bf ESPaDOnS &$\odot$ {\tiny (5)}         \\ 
\bf  9655167 & 5036 & 109 & 3.03 & 0.25 &$-$0.01 & 0.21 &  4.5 & 0.5 & G8III   &  5325 &  80 & 3.57 & 0.15 &    0.06 & 0.07 &  1.57 &  0.06 &\bf ESPaDOnS &$\odot$ {\tiny (5)}         \\ 
     9693187 &  --- & --- &  --- &  --- &    --- &  --- &  --- & --- & ---     &  ---  & --- & ---  & ---  &    ---  & ---  &  ---  &  ---  &    ESPaDOnS &$\odot$\,SB2 {\tiny (1,15)} \\ 
\bf  9700679 & 5176 & 158 & 3.37 & 0.32 &   0.04 & 0.22 &  3.7 & 0.9 & G8III   &  5101 &  73 & 3.05 & 0.13 & $-$0.08 & 0.06 &  1.01 &  0.04 &\bf HERMES   & hybrid {\tiny (6)}         \\ 
\bf  9702369 & 5956 & 132 & 4.04 & 0.22 &$-$0.11 & 0.23 &  5.1 & 1.4 & F9IV-V  &  6441 &  78 & 4.54 & 0.11 &    0.14 & 0.06 &  1.39 &  0.05 &\bf HERMES   &$\odot$ {\tiny (1)}         \\ 
\bf  9715099 & 6180 &  93 & 4.07 & 0.21 &   0.07 & 0.22 & 25.1 & 1.2 & F6IV    &  ---  & --- & ---  & ---  &    ---  & ---  &  ---  &  ---  &\bf FRESCO   &$\odot$ {\tiny (1)}         \\ 
\bf  9716090 & 5053 & 106 & 3.17 & 0.24 &   0.02 & 0.21 &  3.3 & 0.6 & G8III   &  5297 &  74 & 3.41 & 0.12 & $-$0.04 & 0.06 &  1.75 &  0.05 &\bf ESPaDOnS &$\odot$ {\tiny (5)}         \\ 
\bf  9716522 & 4860 &  92 & 2.82 & 0.21 &$-$0.03 & 0.21 &  2.7 & 0.3 & G9III   &  5126 &  73 & 3.10 & 0.12 &    0.05 & 0.06 &  1.67 &  0.04 &\bf ESPaDOnS &$\odot$ {\tiny (5)}         \\ 
     9812850 & 6258 &  97 & 3.94 & 0.21 &$-$0.22 & 0.21 &  9.8 & 0.7 & F6IV    &  6790 & 118 & 4.92 & 0.13 & $-$0.04 & 0.08 &  2.70 &  0.27 &    ESPaDOnS &$\odot$ {\tiny (1)}         \\ 
     9908400 & 6068 & 106 & 3.95 & 0.22 &   0.17 & 0.21 & 17.9 & 0.9 & G0IV    &  ---  & --- & ---  & ---  &    ---  & ---  &  ---  &  ---  &    NARVAL   &$\odot$ {\tiny (1)}         \\ 
     9955598 & 5264 &  95 & 4.29 & 0.22 &$-$0.04 & 0.21 &  1.2 & 0.6 & K0V     &  5380 &  68 & 4.33 & 0.12 &    0.04 & 0.06 &  0.80 &  0.06 &    NARVAL   &$\odot\,\wp?$ {\tiny (1,10)}\\ 
\bf  9965715 & 6326 & 116 & 4.00 & 0.21 &$-$0.30 & 0.21 &  8.2 & 0.7 & F2V     &  6542 &  87 & 4.71 & 0.12 & $-$0.22 & 0.06 &  1.84 &  0.10 &\bf ESPaDOnS                              \\ 
\bf 10001154 & 4391 &  96 & 2.17 & 0.22 &$-$0.23 & 0.21 &  2.6 & 0.2 & G9III   &  4585 &  82 & 2.34 & 0.16 & $-$0.20 & 0.06 &  2.06 &  0.06 &\bf ESPaDOnS                              \\ 
\bf 10010623 & 6464 & 106 & 4.11 & 0.21 &$-$0.01 & 0.21 & 31.8 & 2.1 & F3V     &  ---  & --- & ---  & ---  &    ---  & ---  &  ---  &  ---  &\bf ESPaDOnS                              \\ 
    10016239 & 6214 & 103 & 3.95 & 0.21 &$-$0.17 & 0.21 & 10.7 & 1.0 & F6IV    &  ---  & --- & ---  & ---  &    ---  & ---  &  ---  &  ---  &    NARVAL   &$\odot$ {\tiny (1)}         \\ 
    10018963 & 6145 & 112 & 3.95 & 0.21 &$-$0.27 & 0.21 &  2.1 & 0.6 & F6IV    &  6354 &  69 & 4.32 & 0.11 & $-$0.16 & 0.05 &  1.79 &  0.05 &    NARVAL   &$\odot$ {\tiny (1)}         \\ 
    10068307 & 6144 & 109 & 3.94 & 0.21 &$-$0.22 & 0.21 &  3.4 & 0.8 & F6IV    &  6288 &  68 & 4.28 & 0.10 & $-$0.11 & 0.06 &  1.68 &  0.04 &    ESPaDOnS &$\odot$ {\tiny (1)}         \\ 
\bf 10079226 & 5854 &  97 & 4.27 & 0.21 &   0.10 & 0.21 &  1.6 & 1.2 & G0V     &  6045 &  68 & 4.49 & 0.11 &    0.17 & 0.06 &  1.17 &  0.04 &\bf HERMES   &$\odot$ {\tiny (1)}         \\ 
    10124866 & 5736 &  92 & 4.29 & 0.21 &$-$0.31 & 0.21 &  3.0 & 0.6 & G4V     &  5864 &  68 & 4.57 & 0.11 & $-$0.24 & 0.06 &  1.03 &  0.05 &    ESPaDOnS &$\odot$ {\tiny (9)}         \\ 
\bf 10131030 & 4897 &  89 & 2.74 & 0.21 &   0.02 & 0.21 &  3.0 & 0.9 & G8III   &  ---  & --- & ---  & ---  &    ---  & ---  &  ---  &  ---  &\bf FRESCO                                \\ 
    10162436 & 6149 & 115 & 3.95 & 0.21 &$-$0.16 & 0.22 &  2.8 & 0.8 & F8IV    &  6423 &  71 & 4.43 & 0.11 &    0.01 & 0.06 &  1.75 &  0.05 &    ESPaDOnS &$\odot$ {\tiny (1)}         \\ 
    10355856 & 6351 & 118 & 3.93 & 0.21 &$-$0.22 & 0.21 &  4.5 & 0.8 & F5IV-V  &  6612 &  79 & 4.38 & 0.11 & $-$0.01 & 0.06 &  1.84 &  0.05 &    ESPaDOnS &$\odot$ {\tiny (1)}         \\ 
\bf 10388249 & 4743 &  92 & 2.87 & 0.22 &   0.00 & 0.21 & 10.6 & 0.2 & K1IV    &  4978 &  98 & 3.48 & 0.19 &    0.14 & 0.07 &  1.87 &  0.09 &\bf FIES     &$\odot$ {\tiny (1)}         \\ 
    10454113 & 6129 & 151 & 4.07 & 0.22 &$-$0.16 & 0.22 &  3.7 & 1.0 & F9IV-V  &  6216 &  68 & 4.46 & 0.10 &    0.00 & 0.05 &  1.30 &  0.04 &    ESPaDOnS &$\odot$ {\tiny (1)}         \\ 
    10462940 & 6026 & 101 & 4.24 & 0.21 &   0.05 & 0.21 &  1.9 & 0.7 & G0.5IV  &  6268 &  68 & 4.48 & 0.10 &    0.18 & 0.05 &  1.35 &  0.03 &    NARVAL   &$\odot$ {\tiny (1)}         \\ 
    10516096 & 5928 &  95 & 4.24 & 0.21 &$-$0.04 & 0.21 &  2.8 & 0.6 & F9IV-V  &  6094 &  70 & 4.47 & 0.11 & $-$0.03 & 0.06 &  1.39 &  0.05 &    ESPaDOnS &$\odot$ {\tiny (1)}         \\ 
\bf 10526137 & 3316 & 274 & 3.93 & 0.57 &$-$0.23 & 0.21 & 13.4 & 1.6 & M2V     &  ---  & --- & ---  & ---  &    ---  & ---  &  ---  &  ---  &\bf FIES     &APER {\tiny (2)}            \\ 
    10644253 & 5910 &  93 & 4.30 & 0.21 &   0.05 & 0.21 &  1.6 & 0.7 & G0V     &  6132 &  65 & 4.54 & 0.11 &    0.15 & 0.05 &  1.21 &  0.03 &    ESPaDOnS &$\odot$ {\tiny (1)}         \\ 
    10709834 & 6398 & 124 & 3.94 & 0.21 &$-$0.20 & 0.21 &  7.0 & 1.2 & F5IV-V  &  ---  & --- & ---  & ---  &    ---  & ---  &  ---  &  ---  &    NARVAL   &$\odot$ {\tiny (1)}         \\ 
\bf 10735274 & 3836 & 202 & 1.72 & 0.36 &$-$0.06 & 0.22 &  2.8 & 1.5 & K5III   &  ---  & --- & ---  & ---  &    ---  & ---  &  ---  &  ---  &\bf HERMES                                \\ 
    ...~~~~~ & 4033 &  99 & 1.69 & 0.23 &$-$0.17 & 0.20 &  9.3 & 2.0 & K4III   &  ---  & --- & ---  & ---  &    ---  & ---  &  ---  &  ---  &\bf FRESCO                                \\ 
    10923629 & 6109 &  99 & 4.00 & 0.21 &   0.08 & 0.21 &  7.3 & 0.8 & F8V     &  ---  & --- & ---  & ---  &    ---  & ---  &  ---  &  ---  &    NARVAL   &$\odot$ {\tiny (1)}         \\ 
    10963065 & 6097 & 130 & 4.00 & 0.21 &$-$0.27 & 0.22 &  2.3 &  0.6& F8V     &  6236 &  64 & 4.55 & 0.11 & $-$0.15 & 0.05 &  1.47 &  0.03 &    NARVAL   &$\odot\,\wp?$ {\tiny (1,10)}\\ 
\bf 11018874 & 6454 & 121 & 4.08 & 0.21 &$-$0.04 & 0.21 & 49.0 & 2.3 & F5V     &  ---  & --- & ---  & ---  &    ---  & ---  &  ---  &  ---  &\bf ESPaDOnS                              \\ 
    11026764 & 5771 &  97 & 4.22 & 0.21 &   0.10 & 0.21 &  2.6 & 0.9 & G1V     &  5802 &  68 & 4.12 & 0.11 &    0.11 & 0.06 &  1.30 &  0.04 &    ESPaDOnS &$\odot$ {\tiny (1)}         \\ 
\bf 11037105 & 6801 & 132 & 4.20 & 0.22 &$-$0.14 & 0.23 & 27.9 & 2.0 & F2V     &  ---  & --- & ---  & ---  &    ---  & ---  &  ---  &  ---  &\bf ESPaDOnS                              \\ 
    11081729 & 6400 & 127 & 3.97 & 0.21 &$-$0.19 & 0.22 & 21.4 & 0.7 & F5IV    &  ---  & --- & ---  & ---  &    ---  & ---  &  ---  &  ---  &    ESPaDOnS &$\odot$ {\tiny (1)}         \\ 
\bf 11099165 & 3930 &  90 & 1.69 & 0.21 &$-$0.18 & 0.21 &  2.5 & 0.7 & K5III   &  ---  & --- & ---  & ---  &    ---  & ---  &  ---  &  ---  &\bf ESPaDOnS                              \\ 
    11137075 & 5576 &  99 & 4.14 & 0.22 &$-$0.04 & 0.21 &  2.3 & 0.4 & G5IV-V  &  5610 &  71 & 4.10 & 0.12 & $-$0.06 & 0.06 &  1.10 &  0.04 &    NARVAL   &$\odot$ {\tiny (1)}         \\ 
    11244118 & 5605 & 104 & 4.05 & 0.23 &   0.19 & 0.21 &  1.7 & 0.5 & G5IV    &  5770 &  67 & 4.14 & 0.11 &    0.35 & 0.06 &  1.19 &  0.03 &    NARVAL   &$\odot$ {\tiny (1)}         \\ 
    11253226 & 6410 & 125 & 3.96 & 0.21 &$-$0.20 & 0.21 & 11.4 & 1.2 & F5IV-V  &  ---  & --- & ---  & ---  &    ---  & ---  &  ---  &  ---  &    ESPaDOnS &$\odot$ {\tiny (1)}         \\ 
\hline                                     
\end{tabular}                              
}
\end{center}
\end{table}
\end{landscape}

\setcounter{table}{2}
\begin{landscape}
\begin{table}
\begin{center}
\caption{continuation.}
{\small
\begin{tabular}{rrrrrrrrrlrrrrrrrrll}
\hline\hline\noalign{\smallskip}
KIC\hspace{16pt}     & $T_{\rm eff}$ & $\sigma$ & $\log g$ & $\sigma$ & [Fe/H] & $\sigma$ & $v\sin i$ & $\sigma$ & MK
                     & $T_{\rm eff}$ & $\sigma$ & $\log g$ & $\sigma$ & [Fe/H] & $\sigma$ & $\xi_{\rm t}$ & $\sigma$ & Instrument & var\\
         &\multicolumn{9}{c}{\hrulefill \,ROTFIT\,\hrulefill}
         &\multicolumn{8}{c}{\hrulefill \,ARES+MOOG\,\hrulefill} &\\

\hline
\hline
\bf 11342410 & 5858 & 110 & 4.26 & 0.21 &$-$0.07 & 0.22 &  1.8 & 0.6 & G1V     &  ---  & --- & ---  & ---  &    ---  & ---  &  ---  &  ---  &\bf FRESCO   \\                    
\bf 11396108 & 6330 & 169 & 3.97 & 0.22 &$-$0.03 & 0.21 & 20.1 & 1.9 & F6IV    &  ---  & --- & ---  & ---  &    ---  & ---  &  ---  &  ---  &\bf FRESCO   \\                    
    11414712 & 5731 &  93 & 4.16 & 0.21 &   0.02 & 0.21 &  2.3 & 0.9 & G3V     &  5725 &  61 & 3.99 & 0.10 & $-$0.02 & 0.05 &  1.27 &  0.01 &    NARVAL   &$\odot$ {\tiny (1)}\\
\bf 11495120 & 4864 &  90 & 2.70 & 0.21 &$-$0.09 & 0.21 &  2.9 & 0.5 & G8III   &  ---  & --- & ---  & ---  &    ---  & ---  &  ---  &  ---  &\bf FRESCO   &$\odot$ {\tiny (4)}\\
    11498538 & 6453 & 123 & 4.07 & 0.22 &$-$0.01 & 0.21 & 33.2 & 1.3 & F2V     &  ---  & --- & ---  & ---  &    ---  & ---  &  ---  &  ---  &    ESPaDOnS &rot/act {\tiny (6)}\\
\bf 11551430 & 5649 & 141 & 4.01 & 0.22 &$-$0.07 & 0.22 & 24.3 & 0.5 & G5IV    &  ---  & --- & ---  & ---  &    ---  & ---  &  ---  &  ---  &\bf FRESCO   \\                    
\bf 11559263 & 5633 & 175 & 4.02 & 0.26 &   0.08 & 0.21 &  5.3 & 0.6 & G5III   &  5284 &  66 & 3.03 & 0.11 & $-$0.02 & 0.06 &  0.77 &  0.03 &\bf HERMES   \\                    
\bf 11708170 & 6872 & 124 & 4.21 & 0.22 &$-$0.04 & 0.21 & 32.9 & 2.3 & F1V     &  ---  & --- & ---  & ---  &    ---  & ---  &  ---  &  ---  &\bf ESPaDOnS &rot/act {\tiny (6)}\\
\bf 11709006 & 5852 & 104 & 4.38 & 0.21 &   0.01 & 0.21 & 10.2 & 0.2 & G1.5V   &  6047 &  79 & 4.66 & 0.11 &    0.05 & 0.06 &  1.40 &  0.07 &\bf HERMES   \\                    
    11717120 & 5155 & 104 & 3.76 & 0.27 &$-$0.17 & 0.21 &  0.6 & 0.3 & G9.5IV  &  5118 &  67 & 3.80 & 0.12 & $-$0.27 & 0.06 &  0.89 &  0.04 &\bf FIES     &$\odot$ {\tiny (1)}\\
    ...~~~~~ & 5222 & 109 & 3.82 & 0.24 &$-$0.17 & 0.21 &  1.1 & 0.4 & G8IV    &  5137 &  65 & 3.87 & 0.12 & $-$0.28 & 0.05 &  0.83 &  0.04 &    NARVAL   \\                    
    11179629 &  --- & --- &  --- &  --- &    --- &  --- &  --- & --- & ---     &  ---  & --- & ---  & ---  &    ---  & ---  &  ---  &  ---  &    ESPaDOnS &SB2 {\tiny (15)}\\   
\bf 11754082 & 4742 &  94 & 2.77 & 0.23 &$-$0.10 & 0.21 & 11.5 & 2.7 & G9III   &  ---  & --- & ---  & ---  &    ---  & ---  &  ---  &  ---  &\bf FRESCO   &$\odot$ {\tiny (4)}\\
\bf 11772920 & 5209 & 121 & 4.34 & 0.23 &$-$0.07 & 0.21 &  1.4 & 0.8 & K1V     &  5341 &  80 & 4.44 & 0.13 & $-$0.10 & 0.06 &  0.73 &  0.10 &\bf HERMES   &$\odot$ {\tiny (9)}\\
    12009504 & 6099 & 125 & 4.00 & 0.21 &$-$0.14 & 0.22 &  5.9 & 0.7 & F9IV-V  &  6267 &  71 & 4.37 & 0.11 & $-$0.03 & 0.06 &  1.59 &  0.06 &    ESPaDOnS &$\odot$ {\tiny (1)}\\
\bf 12155015 & 3937 &  91 & 1.68 & 0.21 &$-$0.16 & 0.21 &  2.8 & 0.9 & K5III   &  ---  & --- & ---  & ---  &    ---  & ---  &  ---  &  ---  &\bf ESPaDOnS \\                    
    12258514 & 5952 &  94 & 4.23 & 0.21 &   0.06 & 0.21 &  1.7 & 0.6 & G0.5IV  &  6099 &  66 & 4.32 & 0.10 &    0.10 & 0.05 &  1.36 &  0.03 &    ESPaDOnS &$\odot$ {\tiny (1)}\\
\bf 12453925 & 6514 & 153 & 4.14 & 0.22 &$-$0.02 & 0.21 & 75.2 & 2.9 & F3V     &  ---  & --- & ---  & ---  &    ---  & ---  &  ---  &  ---  &\bf ESPaDOnS \\                    
    12455203 & 4919 &  93 & 2.89 & 0.22 &$-$0.02 & 0.21 &  2.3 & 0.3 & G8III   &  5104 &  69 & 3.14 & 0.13 &    0.07 & 0.06 &  1.49 &  0.03 &    ESPaDOnS &$\odot$ {\tiny (4)}\\
\bf 12508433 & 5134 & 121 & 3.50 & 0.28 &   0.08 & 0.22 &  0.6 & 0.4 & K0III-IV&  5281 &  76 & 3.85 & 0.13 &    0.21 & 0.06 &  0.98 &  0.06 &\bf HERMES   &$\odot$ {\tiny (1)}\\
\hline
\end{tabular}
}
\begin{minipage}{23cm}
{\tiny $^*$ We use an asterisk to indicate stars listed by \citet{bruntt2012} or {\citet{thygesen2012} who do not provide the information whether they used the
ESPaDOnS or the NARVAL spectra  in their analysis.\\ 

The equatorial coordinates and multi-colour magnitudes of the stars from this table are available at the Mikulski Archive for Space Telescopes (MAST) at 
\url{http://archive.stsci.edu/kepler/kepler_fov/search.php}} \vspace{5pt}

Literature:\\
(1) \citet{huber2011, chaplin2011}, (2) \citet{pigulski2009}, (3) \citet{howell2012}, (4) \citet{hekker2011}, (5) \citet{stello2011}, (6) \citet{uytterhoeven2011}, 
(7) \citet{sergey2007}, (8) \citet{smelcer2003}, (9) \citet{appourchaux2012}, (10) \url{http://planetquest.jpl.nasa.gov/kepler/table}, (11) \citet{frohlich2012}, 
(12) \citet{frasca2011}, (13) \citet{verner2011}, (14) Thygesen et al. (in preparation), (15) this paper. \vspace{5pt}

Types of variability:\\
$\odot$: solar-like oscillations,
$\delta$ Sct: $\delta$~Scuti--type pulsations,
Mira: $o$~Ceti--type pulsations,
RS\,CVn: RS\,CVn--type variability,
rot/act: star activity/rotational modulation,
PER: strictly periodic with sinusoidal light-curves,
APER: no well-pronounced periodicity,
$\wp$: planet-hosting star,
$\wp?$: candidate for a planet-hosting star,
SB2: double-lined spectroscopic binary.

}

\end{minipage}
\end{center}
\end{table}
\end{landscape}

\label{lastpage}

\end{document}